\documentclass[12pt]{article}

\usepackage{fancyhdr,titlesec,geometry,ragged2e}
\usepackage[dvipsnames]{xcolor}
\usepackage[many]{tcolorbox}
\usepackage{layout}
\usepackage{lipsum}
\usepackage{hyperref}
\hypersetup{
  colorlinks,
  citecolor=red,
  linkcolor=blue,
  linktoc=all
}

\usepackage[T1]{fontenc}                            
\usepackage{lmodern,mathrsfs}

\usepackage[shortlabels]{enumitem}
\usepackage{mathtools,amssymb,amsfonts,amsthm,bm,mathrsfs}   
\usepackage{array,tabularx,booktabs}                
\usepackage{graphicx,wrapfig,float,caption,epsfig}         
\usepackage{setspace,multicol}                      
\usepackage{tikz,tikz-cd,physics,cancel}            
\usepackage{esint}
\allowdisplaybreaks

\usepackage{parskip}
\DeclareGraphicsExtensions{.pdf,.png,.jpg}
\usepackage{geometry}
\renewcommand{\thefootnote}{\arabic{footnote}}

\titleformat{\section}[hang]{\normalfont\large\bfseries}{\thesection.}{3mm}{}
\titlespacing{\section}{0mm}{15mm}{4mm}

\titleformat{\subsection}[hang]{\normalfont\bfseries}{\thesubsection}{2mm}{}
\titlespacing{\subsection}{0mm}{10mm}{4mm}

\titleformat{\subsubsection}[hang]{\normalfont\bfseries}{\thesubsubsection}{5mm}{}
\titlespacing{\subsubsection}{0mm}{5mm}{1mm}

\newtheoremstyle{thmstyle}
{7pt} 
{3pt} 
{\itshape} 
{} 
{\bfseries} 
{.} 
{0.5em} 
{} 

\theoremstyle{remark}{

}

\theoremstyle{thmstyle}{
\newtheorem{definition}{Definition}[section]
\newtheorem{theorem}{Theorem}[section]
\newtheorem{proposition}{Proposition}[section]
\newtheorem{lemma}{Lemma}[section]
\newtheorem{corollary}{Corollary}[theorem]
}

\numberwithin{equation}{section}
\renewcommand{\theequation}{\thesection.\arabic{equation}}
\newlength{\extraspace}
\newlength{\extraspaces}
\newcommand{\be}{\begin{equation}
\addtolength{\abovedisplayskip}{\extraspaces}
\addtolength{\belowdisplayskip}{\extraspaces}
\addtolength{\abovedisplayshortskip}{\extraspace}
\addtolength{\belowdisplayshortskip}{\extraspace}}
\newcommand{\ee}{\end{equation}}
\newcommand{\ba}{\begin{eqnarray}
\addtolength{\abovedisplayskip}{\extraspaces}
\addtolength{\belowdisplayskip}{\extraspaces}
\addtolength{\abovedisplayshortskip}{\extraspace}
\addtolength{\belowdisplayshortskip}{\extraspace}}
\newcommand{\ea}{\end{eqnarray}}
\newcommand{\bas}{\begin{eqnarray*}
\addtolength{\abovedisplayskip}{\extraspaces}
\addtolength{\belowdisplayskip}{\extraspaces}
\addtolength{\abovedisplayshortskip}{\extraspace}
\addtolength{\belowdisplayshortskip}{\extraspace}}
\newcommand{\eas}{\end{eqnarray*}}

\newcounter{subequation}[equation]
\makeatletter

\expandafter\let\expandafter
\reset@font\csname reset@font\endcsname

\def\subeqnarray{\arraycolsep1pt
    \def\@eqnnum\stepcounter##1{\stepcounter{subequation}%
        {\reset@font\rm(\theequation\alph{subequation})}}
\jot5mm     \eqnarray}

\def\subarray{\arraycolsep1pt
    \def\@eqnnum\stepcounter##1{\stepcounter{subequation}%
        {\reset@font\rm(\alph{subequation})}}
\jot5mm     \eqnarray}

\makeatother

\newcommand{\pp}{{\partial}}
\renewcommand{\dd}{{\partial}}

\newcommand{\half}{\textstyle{1\over 2}}

\newcommand{\ssum}[2]{\sum_{#1}^{#2}}
\newcommand{\exx}[1]{e^{#1}}

\newcommand{\ra}{\rightarrow}

\newcommand{\sspace}{\makebox[1cm]{ }}

\newcommand{\nspace}{\!\!\!\!\!\!\!\!\!\!}

\newcommand{\nonum}{\nonumber \\[1.5mm]}
\newcommand{\is }{&\!\!=\!\!&} 
\newcommand{\isa}{ & = } 

\DeclarePairedDelimiterX\set[1]\lbrace\rbrace{#1}

\newcommand{\inv}{^{-1}}

\newcommand{\lb}{\lambda}

\newcommand{\om}{\omega}

\newcommand{\al}{\alpha}

\newcommand{\vp}{\varphi}

\newcommand{\munu}{{\mu\nu}}

\newcommand{\mcD}{{\mathcal{D}}}
\newcommand{\mcE}{{\mathcal{E}}}

\newcommand{\mcK}{{\mathcal{K}}}

\newcommand{\cD}{{\cal D}}
\newcommand{\cE}{{\cal E}}

\newcommand{\cS}{{\cal S}}

\newcommand{\bbC}{{\mathbb{C}}}

\newcommand{\bbH}{{\mathbb{H}}}

\newcommand{\bbN}{{\mathbb{N}}}

\newcommand{\bbR}{{\mathbb{R}}}
\newcommand{\R}{\mathbb{R}}
\newcommand{\bbS}{{\mathbb{S}}}

\newcommand{\mfb}{{\mathfrak{b}}}

\newcommand{\wg}[0]{{\rm g}}

\renewcommand{\theequation}{\thesection.\arabic{equation}}
\numberwithin{equation}{section}

\def\fnum@figure{\textbf{\figurename\nobreakspace\thefigure}}
\def\fnum@table{\textbf{\tablename\nobreakspace\thetable}}

\newcommand{\FL}[0]{Friedmann-Lema\^{i}tre }

\newcommand{\mdp}[0]{|p|}

\newcommand{\itw}[1]{W^{(#1)}}
\newcommand{\vacw}[1]{S^{(#1)}}

\newcommand{\bsle}[0]{$\mfb$SLE$\;$}

\newcommand{\scrA}[1]{\mathscr{A}_{#1}(J)}

\begin{document}

\begin{titlepage}

\renewcommand{\thefootnote}{\arabic{footnote}}
\makebox[1cm]{}
\vspace{1cm}

\begin{center}
\mbox{{\Large \bf States of Low Energy on Bianchi I spacetimes}} 
\vspace{2.8cm}
\hypersetup{linkcolor=black}

{\sc R.~Banerjee}\footnote[1]{email: \ttfamily\href{mailto:rudrajit.banerjee@oist.jp}{\textcolor{black}{rudrajit.banerjee@oist.jp}}} 
{\sc and} 
{\sc M.~Niedermaier}\footnote[2]{email: \ttfamily\href{mailto:mnie@pitt.edu}{\textcolor{black}{mnie@pitt.edu}}}
\\[8mm]
{\small\sl $^1$Okinawa Institute of Science and Technology Graduate University,}\\
{\small\sl 1919-1, Tancha, Onna, Kunigami District,}\\
{\small\sl Okinawa 904-0495, Japan}
\\[5mm]
{\small\sl $^2$Department of Physics and Astronomy}\\
{\small\sl University of Pittsburgh, 100 Allen Hall}\\
{\small\sl Pittsburgh, PA 15260, USA}
\vspace{18mm}

{\bf Abstract} \\[4mm]

\begin{quote}
States of Low Energy are a class of exact Hadamard states for free quantum fields
on cosmological spacetimes whose structure is fixed at {\it all} scales by a minimization 
principle. The original construction was for \FL geometries and is here generalized 
to anisotropic Bianchi I geometries relevant to primordial cosmology. In addition to proving 
the Hadamard property, systematic series expansions in the infrared and ultraviolet 
are developed. The infrared expansion is convergent and induces in the massless case
a leading spatial long distance decay that is always Minkowski-like but anisotropy modulated. The ultraviolet expansion is shown to be equivalent to the  Hadamard property, and a non-recursive formula for its coefficients is presented.      
\end{quote}
\end{center}
\vfill
\setcounter{footnote}{0}
\end{titlepage}
\thispagestyle{empty}
\makebox[1cm]{}

\vspace{-23mm}
\begin{samepage}
{\hypersetup{linkcolor=black}
\tableofcontents}
\end{samepage}

\nopagebreak

\newpage



\section{Introduction}

Hadamard states formalize the notion of a physically acceptable free state 
for perturbatively defined quantum field theories on globally hyperbolic 
spacetimes. Once the free state on which perturbation theory is based satisfies 
the Hadamard condition, all Wick powers of the field are well-defined and 
the perturbative series exists termwise to any order, see 
\cite{KhavMoretti,HollWaldRept,Gerardbook} for recent surveys. Although general results  
ensure the existence of Hadamard states and the associated two-point functions
on generic globally hyperbolic backgrounds \cite{FNW,Gerardbook}, any computational control beyond 
an ultraviolet asymptotic expansion requires a more concrete construction. Such 
constructions are surprisingly difficult
\cite{HadamardNullbook,JunkerS,SJstates,SJHadamard} even for background 
spacetimes with some degree of symmetry, other than maximal.

On generic \FL spacetimes the so-called States of Low Energy (SLEs), introduced by Olbermann 
\cite{Olbermann} provide a concrete construction of exact Hadamard states.  
These may be regarded as an improvement of the often pathological \cite{FRWHamdiag0} 
instantaneous Hamiltonian diagonalization procedure, with the SLE arising by minimizing 
the Hamiltonian's expectation value {\it after} averaging with a temporal window function 
$f$. In addition to being exact Hadamard states SLEs on \FL spacetimes have a number of 
further desirable properties \cite{BonusSLE}. First, an SLE is fully 
determined by the commutator function and thus is unique for fixed window function.
Second, it admits a convergent expansion in powers of the spatial momentum-modulus. For massless 
theories the leading infrared behavior is Minkowski-like for all scale factors
and all window functions.     
Third, the general Hadamard expansion can be replaced with a considerably simpler 
one in inverse powers of the spatial momentum-modulus. Its coefficients can be obtained by 
an easily programmable recursion and are manifestly independent of the window function. 

Several generalizations and applications have been considered. By analyzing massless 
SLEs for suitable interpolating line elements they were found to be viable 
vacua for cosmological perturbations in a pre-inflationary period
\cite{BonusSLE}. A related theme is the extension to bounce cosmologies with the 
window function overlapping with the bounce \cite{LQCSLE1,LQCSLE2,CPTSLE}. 
In the context of loop quantum cosmology the effective modifications can be
attributed to a time dependent mass and the robustness of the ensued power spectrum 
has been investigated \cite{LQCSLE1,LQCSLE2}. A radiation dominated 
background was interpreted as a bounce in \cite{CPTSLE} with CPT invariant 
quantum fields in SLE-type vacua placed on  it, including an extension 
to fermions. An application to the Schwinger effect can be found 
in \cite{SLESchwinger} with the focus on the directional dependence induced by 
the external electric field.  Finally, the  SLE construction has been generalized 
in \cite{SLE2} to backgrounds with line element $ds^2=-dt^2+a(t)^2 h_{ij}(x)dx^idx^j$. 
However, in four dimension these do not produce new perfect 
fluid solutions to the Einstein field equations \cite{genFL}.

In this paper, an analysis of SLEs on Bianchi I spacetimes with line element 
\begin{equation} 
\label{Bmetric}
ds^2=-N(t)^2dt^2+\wg_{ij}(t)dx^i dx^j\,,
\end{equation}
is presented. Here $\wg_{ij}(t)$, $i,j =1,\ldots, d$, is a merely time dependent spatial 
metric on leaves that are topologically $\bbR^d$. The main characteristics 
of Bianchi geometries is the existence of a $d$ dimensional Lie group 
of isometries acting transitively on the leaves of constant time. 
For (\ref{Bmetric}) the group is abelian, acts by translations,
and accounts for spatial homogeneity in the familiar sense. 
Often $\wg_{ij}(t)$ is assumed to be diagonal as this is consistent 
with the Einstein equations and a diagonal energy-momentum tensor
\cite{Visserultraloc}. Most of our considerations will be off-shell
and diagonality is not assumed. Metrics of the form 
(\ref{Bmetric}) give rise to important cosmological solutions, which are
especially relevant in the context of the Belinskii, Khalatnikov, 
Lifshitz (BKL) scenario, see \cite{BKLbook, WKB1}. In particular, 
minimally coupled to a homogeneous selfinteracting scalar field, 
Bianchi I spacetimes become asymptotically velocity dominated 
(Kasner-like) towards the Big Bang \cite{BKLBianchi}. At late(r) 
times they typically become isotropic, consistent with the standard 
cosmological paradigm \cite{Nungesser}.  

The extension of the SLE's Hadamard property is less obvious than it may seem.
For example, in an early investigation Fulling \cite{FRWHamdiag0} found that without 
temporal averaging Bianchi I backgrounds lead to infinite particle production 
(incompatible with the Hadamard property) in otherwise identical settings 
where \FL backgrounds do not. 
Likewise, a naive transfer of the small and large momentum expansions mentioned before 
is hampered by the fact that the spatial momentum square $\wg^{ij}(t) p_i p_j$ 
is time dependent. Nevertheless, suitably modified such expansions can still be constructed.  
In overview, our main results are:
\begin{itemize} 
\item [(a)] States of Low Energy can be defined on generic Bianchi I geometries and 
are exact (anisotropic) Hadamard states.
\item [(b)] The associated two point function admits a convergent expansion in powers of 
$|p| = \sqrt{\delta^{ij}p_ip_j}$. In the massless case the leading term is always 
$O(1/|p|)$ with anisotropic modulations determined by the geometry and the window function.  
\item [(c)] The associated two-point function admits an asymptotic UV expansion
whose coefficients are independent of the window function. The expansion 
is shown to be equivalent to the Hadamard property, and   
a non-recursive formula for the coefficients is presented.
\end{itemize} 

The paper is organized as follows. In Section \ref{sec2} we summarize the straightforward 
extension of the SLE construction principle to Bianchi I geometries, including 
an alternative expression solely in terms of the commutator function. Section \ref{sec3}
is devoted to the proof of the Hadamard property. The infrared and ultraviolet 
expansions are developed in Sections \ref{sec4} and \ref{sec5}, respectively. The proofs in
Sections \ref{sec3}--\ref{sec5} draw heavily on properties of adiabatic vacuum states, which 
are prepared in Appendix \ref{AppA}.  


\section{SLE on Bianchi I: construction and basic properties} \label{sec2}

We focus on the  background geometry of a $1\!+\!d$ dimensional Bianchi I spacetime, diffeomorphic to $\bbR_+\times \bbR^d$,  with line element
\begin{eqnarray}\label{metric1}
	ds^2=-N(t)^2 dt^2 + \wg_{ij}(t)dx^i dx^j\,,
\end{eqnarray}
where $N:\bbR_+\to \bbR_+$  is the lapse function. Further, $\wg_{ij}$ is the (time dependent) Riemannian metric on the  spatial slices, for which $x^1,\ldots,x^d$ are adapted coordinates. In these coordinates translations on the spatial sections $\bbR^d$ constitute a transitive abelian group of isometries. The above shift $N^i\equiv 0$ form of the line element (\ref{metric1}) is preserved under ${\rm Diff}[t_i,t_f] \times {\rm ISO}(d)$ transformations, where ${\rm Diff}[t_i,t_f]$ are endpoint preserving reparameterizations of some time interval $[t_i,t_f]$, $0 < t_i < t_f < \infty$, and the Euclidean group ${\rm ISO}(d)$ acts via global spatial diffeomorphisms connected to the identity. Here and later on it will be convenient to use the densitized lapse $n(t):= N(t)/\wg(t)^{1/2}$, with $\wg(t)$ the determinant of the spatial metric components. Throughout we assume $N$ and
$\wg_{ij}$ to be smooth on $\R_+$, with $\wg$ bounded away from zero on compact intervals.

On this background we consider a  scalar field $\chi:\bbR_+\times \bbR^d\to \bbR$, which is minimally coupled and initially selfinteracting with potential $U(\chi)$. Next, we expand the minimally coupled scalar field action $S[\chi]$ on $[t_i, t_f]\times \mathbb{R}^d$ around a spatially homogeneous background scalar $\varphi(t)$ to quadratic order in the fluctuations $\phi(t,x) := \chi(t,x) - \varphi(t)$. This gives a leading term $S[\varphi]$ (proportional to a spatial volume term) whose field equation is one of the evolution equations for a Bianchi I cosmology. For a $\varphi(t)$ solving it (with prescribed $\wg_{ij}(t)$) the term linear in the $\phi$ action reduces to a boundary term and may be omitted. The quadratic piece reads 
\begin{eqnarray}\label{action1}
\!\!	S_2[\phi]\is \frac{1}{2}\int \! dt \!\int\! d^dx \,
	\Big\{ \frac{1}{n(t)} (\partial_t \phi)^2 - 
	n(t) \wg(t) U''(\varphi) \phi^2 
	- n(t) \wg(t)\wg^{ij}(t) \partial_i \phi 
	\partial_j \phi\Big\}.\quad
\end{eqnarray}
So far, $\varphi$ is for prescribed $\wg_{ij}(t)$ a solution of $\- \partial_t(n^{-1} \partial_t \varphi) + n(t) \wg(t) U'(\varphi) =0$, but $\wg_{ij}(t)$ itself is unconstrained. 
As far as the homogeneous background is concerned one could now augment the missing gravitational dynamics by the other Bianchi I field equations. This would turn $\wg_{ij}(t), \varphi(t)$ into a solution of the Einstein equations and classical backreaction effects would be taken into account in the homogeneous sector. As noted in the introduction (and in 
contrast to the generalized \FL geometries \cite{genFL,SLE2}) there
are indeed many cosmologically relevant Bianchi I solutions. The standard ``Quantum Field Theory (QFT) on curved background'' viewpoint, on the other hand,  treats the geometry as external, in which case (\ref{action1}) adheres to the minimal coupling principle only if $U''(\varphi) = m^2$ is identified with a constant mass squared. 

The modern approach to the formulation of a (free) QFT associated with (\ref{action1}) proceeds by 
the construction of a local algebra of observables built from operator solutions of 
the linear Klein-Gordon equation $\delta S_2[\phi]/\delta \phi =0$, with the specification
of a ``vacuum'' playing a secondary role. Disentangling the purely algebraic aspects of the theory   
from the details of the state space realization is especially advantageous on curved backgrounds lacking 
timelike Killing vectors, as there are generically infinitely many physically viable vacuum-like states. 
This holds in particular for Bianchi I backgrounds, but their spatial homogeneity simplifies the analysis. 


\subsection{Homogeneous pure quasi-free states}\label{stateSec}

Specifically, a vacuum-like state on a Bianchi I geometry is an instance of a ``homogeneous pure quasifree state''. 
A ``state'' is initially defined algebraically as a positive linear functional $\om$ over the Weyl algebra \cite{KhavMoretti,HollWaldRept,Gerardbook}. 
In the present context a ``state'' can be identified with the set of multi-point functions it gives rise to. 
Then ``quasifree'' means that only the even $n$-point function are nonzero and can be expressed in terms of the two-point function $\varpi(t,x;t',x')$ via Wick's theorem. Being derived from a ``state'' entails certain properties of 
the two-point function that allow its realization in the form $(\Omega, u(t,x)u(t',x') \Omega)$ via the 
Gelfand-Naimark-Segal (GNS) construction, for field operators $u(t,x)$ acting on vectors $\Omega$ in the reconstructed state space.  Then ``pure'' means that $\Omega$ cannot be written as a convex combination of other states. Finally, for the Bianchi I line element \eqref{metric1}, ``homogeneous'' just means ``translation invariant'', i.e. $\varpi(t,x;t',x')$ depends only on $x\!-\!x'$. 

The operators $u(t,x)$ obtained via the GNS construction turn out to coincide with the Heisenberg field operators $\phi(t,x)$ (which are denoted by the same symbol as the classical field, as the latter will no longer occur). The GNS vector $\Omega$ corresponds to a Fock vacuum $|0_T\rangle$, mapped to zero by annihilation operators defined by a mode expansion of the Heisenberg field operator 
\begin{eqnarray}\label{phiexp} 
	&& \phi(t,x) = \int\! \frac{d^dp}{(2\pi)^d} 
	\big[ T_p(t) {\bf a}_T(p) e^{i px} + T_p(t)^* {\bf a}_T^*(p) e^{-i px} \big] \,,
	\nonumber \\[1.5mm]
	&& \big[ {\bf a}_T(p), {\bf a}_T^*(p')] = (2\pi)^d \delta(p-p')\,,
	\quad {\bf a}_T(p) |0_T\rangle =0\,, 
\end{eqnarray}
where $T_p(t)$ is a Wronskian normalized complex solution of the wave equation, i.e.
\begin{eqnarray}\label{wave1}
	&& \big[(n^{-1} \partial_t)^2 + \wg(t) (m^2 + \wg ^{ij}(t)p_i p_j) \big] T_p(t)=0\,,\quad 
	\nonum 
	&&(n^{-1} \partial_t T_p)(t) T_p(t)^* - 
(n^{-1} \partial_t T_p)(t)^* T_p(t)= -i\,.
\end{eqnarray}
The equal time commutation relations $[\phi(t,p), (n^{-1} \partial_t \phi)(t,p')]  = i (2\pi)^d \delta(p + p')$ 
hold by virtue of the above Wronskian condition. Then 
\begin{eqnarray}
	\label{Bcorr1} 
\varpi(t,x;t',x') = \langle 0_T| \phi(t,x) \phi(t',x') |0_T\rangle
= \int\! \frac{d^dp}{(2 \pi)^d} 
\, T_p(t) \,T_p(t')^\ast \, e^{i p (x-x')}\,.
\end{eqnarray}
One sees that modulo phase choices a ``homogeneous pure quasifree'' state is characterized by a choice of 
Wronskian normalized solution $T_p(t)$ of the wave equation or, equivalently, by a choice of Fock vacuum 
$|0_T\rangle$ via (\ref{phiexp}).  However, not all homogeneous pure quasifree states are physically viable, 
with the Hadamard property widely regarded as an additional necessary and sufficient attribute.

{\bf Conventions.}
We briefly comment on our choice of conventions. In 
(\ref{phiexp}) often the ${\bf a}_T^*(p)$ is paired
with $T_p(t)$ not with $T_p(t)^*$. Then the sign 
in the Wronskian normalization condition has to 
be flipped correspondingly. 
More importantly, we seek to preserve temporal reparameterization 
invariance by carrying the lapse-like $n(t) = N(t)/\wg(t)^{1/2}$ 
along. Since in the wave equation $n$ only occurs in the 
combination $n^{-1} \partial_t$, it is convenient to introduce 
a new time function 
\begin{eqnarray}
	\label{taudef} 
\tau := \int^t_{t_0} \! dt' n(t') \,, \quad 
\partial_{\tau} = n(t)^{-1} \partial_t\,,
\end{eqnarray}
for some $t_0$. Note that $\tau(t) = \tau'(t')$ is a scalar under 
reparameterizations $t' = t'(t)$ of the coordinate time $t$, 
and that $d\tau = dt n(t)$, $n(t)^{-1} 
\delta(t,t_1) = \delta(\tau,\tau_1)$ are likewise invariant. We write $\wg_{ij}(\tau)$ for the metric on the spatial sections viewed as a function of $\tau$ rather than $t$,  similarly for its determinant $\wg(\tau)$, as well as $T_p(\tau)$. The defining relations for $T_p(\tau)$ 
then read
\be
\label{wave2} 
\big[ \partial_{\tau}^2 + \omega_p(\tau)^2 ] T_p(\tau) =0\,,\quad 
(\partial_{\tau} T_p \,T_p^*)(\tau) - (\partial_{\tau} T_p^* \,T_p)(\tau) = -i \,,
\ee
with the following ``Bianchi I dispersion relation'' 
\be 
\label{Bdispersion} 
\omega_p(\tau) = \omega^{\mfb}_p(\tau) := \sqrt{\wg(\tau)\big(m^2 + \wg^{ij}(\tau) p_i p_j\big)}\,, \quad m \geq 0\,.
\ee

We remark that the time coordinate $\tau$ corresponds to the $n(t)\equiv 1$ gauge, which is the Bianchi I counterpart of the proper time gauge $\partial_t n(t,x) =0$ often adopted for the evolution of generic foliated spacetimes.  Finally, we note that in general $(n^{-1} \partial_t)^2 = n^{-2}(\partial_t^2 - n^{-1} \partial_t n \partial_t)$, and hence the  first order term in the wave equation \eqref{wave1}  can be removed by the redefinition $T_p(t) = n(t)^{1/2} \chi_p(t)$,
\begin{eqnarray}
	\label{wave3} 
&& \big[ \partial_t^2 + n(t)^2 \omega_p(t)^2 + s(t)\big] \chi_p(t)=0\,,
\nonumber \\[1.5mm]
&& s(t) := \frac{1}{2} \frac{\partial_t^2 n}{ n} - 
\frac{3}{4} \Big( \frac{\partial_t n}{n} \Big)^2\,,
\nonumber \\[1.5mm]
&& \partial_t \chi_p \chi_p^* - (\partial_t \chi_p)^* \chi_p =-i\,.
\end{eqnarray}

Given a Hadamard state $\omega$, of particular interest is the expectation value of the stress energy tensor in this state, $\langle T^\munu\rangle_\om$, which is coherently renormalizeable when $\omega$ has the Hadamard property. Although the energy density associated to  $\langle T^\munu\rangle_\om$ may be arbitrarily negative \cite{Haag1}, it was observed in \cite{Fewster1} that  smearing $T^\munu$ along the worldline of a causal observer yields an expectation value bounded from below. These quantum energy inequalities motivated the original constructions of States of Low Energy (SLE) on Robertson-Walker \cite{Olbermann} and generalized Robertson-Walker \cite{SLE2} backgrounds. Alternatively, the SLE construction may be regarded as an improvement on the instantaneous Hamiltonian diagonalization procedure: SLE arise by minimizing the  Hamiltonian's expectation value {\itshape after} averaging with a temporal window function. Indeed, the temporal averaging is crucial to avoid the pathologies of instantaneous diagonalization. As mentioned, these become more severe for 
Bianchi I geometries \cite{FRWHamdiag0}.


\subsection{Constructing the States of Low Energy}
\label{subsec2b}

The construction of SLEs proceeds by a minimization principle
pointwise for fixed spatial momentum $p=(p_1,\ldots,p_d)$. The momentum dependence 
enters only through the generalized dispersion relation $\om_p(\tau)$, see
(\ref{wave2}). Technically, we can allow $\om_p(\tau)\neq \om^{\mfb}_p(\tau)$ 
to be a differentiable function of $\tau$ with a largely arbitrary momentum 
dependence, only mildly constrained by the requirement that the 
two-point function (\ref{Bcorr1}) is well-defined. We shall do so throughout Sections 
\ref{subsec2b} and \ref{subsec2c} and comment on the type of $\om_p(\tau)$ needed to ensure the 
Hadamard property at the beginning of Section \ref{sec3}. In the following we 
write $S_p$ for some exact solution of \eqref{wave2}, with some 
largely arbitrary $\om_p(\tau)$. The existence of such solutions is ensured 
for compact $\tau$ intervals by general principles.

The construction of a State of Low Energy on a Bianchi 
I geometry (\bsle for short), takes some such fiducial solution $S_p$ 
as a starting point, considers arbitrary Bogoliubov transformations 
of $S_p$, and minimizes the temporal average of the energy density 
(per mode $p$) with respect to them. This results in a new solution, 
$T_p^{\text \bsle}(\tau)$, which up to a constant phase turns out to be
{\it independent} of the $S_p$ used in its construction. Further, the momentum
dependence is automatically transformed into one conducive to the 
Hadamard property.

The expectation value of the Hamilton operator in a Fock vacuum state $|0_S\rangle$ 
associated with a solution $S_p$ is 
\begin{eqnarray}\label{Bsle2}
	\langle 0_S|\hat{\bbH}(\tau)|0_S\rangle \is \int\!\frac{d^dp}{(2\pi)^d} E_p(\tau)\,,\quad E_p(\tau):=\frac{1}{2}\big(|\pp_\tau S_p(\tau)|^2+\omega_p(\tau)^2 |S_p(\tau)|^2\big)\,.\quad 
\end{eqnarray}
While $\langle 0_S|\hat{\bbH}(\tau)|0_S\rangle$ diverges in lieu of regularization, the contribution $E_p(\tau)$ of the individual (spatial) Fourier modes is finite. Averaging with the square of a smooth, real valued  window function $f(\tau)$ of compact support, we define the smeared functionals
\begin{eqnarray}
	\label{HHam3} 
{\cal E}_p[S] \!&\!:=\!& \!\frac{1}{2} \int\! d\tau\, f(\tau)^2 \, 
\Big\{ |\partial_{\tau} S_p|^2 + \omega_p(\tau)^2 |S_p|^2 \Big\} > |{\cal D}_p[S]|\,,
\nonumber \\[1.5mm]
{\cal D}_p[S] \!&\!:=\!&\! \frac{1}{2} \int\! d\tau f(\tau)^2 \,\Big\{ 
(\partial_{\tau} S_p)^2 + \omega_p(\tau)^2 S_p^2 \Big\} \,.
\end{eqnarray} 
Here $\mcE_p[S]> 0$ is the temporally smeared energy density per mode; the significance of $\mcD_p[S]$ will become clear shortly. 

Next, fixing some  fiducial solution $S_p$ of \eqref{wave2}, an arbitrary solution $T_p$ may be reached by a Bogoliubov transformation
\begin{equation}
	\label{sle1}  
	T_p(\tau)=\lambda_p S_p(\tau)+\mu_p S_p(\tau)^\ast\,,\quad |\lambda_p|^2 - |\mu_p|^2 =1\,.
\end{equation}
With $S_p$ and $p$ held fixed the minimization is then over the parameters $\lambda_p,\mu_p \in \mathbb{C}$.  
By multiplying $T_p(\tau)$ with $e^{- i {\rm arg \mu_p}}$ we may take $\mu_p$ to be real. 
Then $|\lambda_p| = \sqrt{1 + \mu_p^2}$
is fixed and only $\mu_p$ and the phase of $\lambda_p$ are real parameters over which the minimum of ${\cal E}_p[T_p]$ is sought. Inserting \eqref{sle1}  with the simplified parameterization into (\ref{HHam3}) one has 
\begin{eqnarray}
	\label{sle2} 
	{\cal E}_p[T] &\! =\! & (1 + 2 \mu_p^2) {\cal E}_p[S] + \mu_p \sqrt{1 + \mu_p^2} \big( e^{i \arg \lambda_p}{\cal D}_p[S] + e^{- i \arg \lambda_p} {\cal D}_p[S]^* \big)\,,
\nonumber \\[1.5mm]
{\cal D}_p[T] &\! =\! & (1 + \mu_p^2)e^{2 i \arg \lambda_p} {\cal D}_p[S] + \mu_p^2 {\cal D}_p[S]^* + 2 \mu_p \sqrt{1 + \mu_p^2} e^{i \arg \lambda_p} {\cal E}_p[S]\,.
\end{eqnarray}
Clearly, the minimizing phase is such that $e^{i \arg \lambda_p} e^{i \arg {\cal D}_p[S]} =-1$.
Subject to this one can verify $\pp {\cal E}_p[T]/\pp \mu_p \propto {\cal D}_p[T]$. A minimizer 
of ${\cal E}_p[T]$ will therefore be a zero of ${\cal D}_p[T]$. The remaining minimization in 
$\mu_p$ then is straightforward and results in 
\begin{align}
\label{sle3} 
\mu_p=\sqrt{\frac{c_1}{2\sqrt{c_1^2-|c_2|^2}}-\frac{1}{2}}\,,
\quad \lambda_p=-\,e^{-i\,{\rm arg} \,c_2}
\sqrt{\frac{c_1}{2\sqrt{c_1^2-|c_2|^2}}+\frac{1}{2}}\,.
\end{align}
Here we adopt a standard notation and set
\begin{eqnarray}
	\label{sle4}
c_1 &:=& {\cal E}_p[S]= \frac{1}{2} \int\! d\tau f(\tau)^2 
\big[ |\partial_{\tau} S_p|^2 + \omega_p^2 |S_p|^2 \big] > |c_2|\,,
\nonumber \\[1.5mm]
c_2 &:=& {\cal D}_p[S]= \frac{1}{2} \int\! d\tau f(\tau)^2 
\big[ (\partial_{\tau} S_p)^2 + \omega_p^2 S_p^2 \big]\,,
\end{eqnarray}
whenever the fiducial solution is clear from the context. 
Since only a phase choice has been made in arriving at (\ref{sle4}) it is clear that the minimizing 
linear combination is unique up to a phase, {\it for a fixed 
fiducial solution $S$.} It is called the {\it Bianchi I State of Low Energy} 
($\mfb$SLE) solution of \eqref{wave2} with fiducial solution $S$. We denote this potential dependence on the fiducial solution $S$ by writing the minimizer as 
\begin{equation}
	\label{sle5} 
T_{p}[S](\tau) := \lambda_p[S] S_p[\tau] + \mu_p[S] S_p(\tau)^*\,,
\end{equation}
with {$\lambda_p[S], \mu_p[S]$} the functionals from (\ref{sle3}), (\ref{sle4}). We now show, however, that the Wightman function associated to this solution (and the associated state) is {\itshape independent of the fiducial solution}:
\begin{theorem}
\label{thbogo}
The \bsle  two-point function based on a fiducial solution $S$
\begin{eqnarray}\label{bsle3}
	\varpi[S](\tau,x;\tau',x'):=\int\!\frac{d^dp}{(2\pi)^d}e^{ip(x-x')}T_{p}[S](\tau)T_{p}[S](\tau')^\ast \,,
\end{eqnarray}
is a Bogoliubov invariant, i.e.~$\varpi[aS+bS^\ast]=\varpi[S]$ with $a,b\in \mathbb{C}$, $|a|^2-|b|^2=1$. Hence the above \bsle defines a homogeneous pure quasi-free state $\om^{\text \bsle}$ independent of the choice of fiducial solution used to realize it.
The associated solution of the wave equation is denoted by $T_p^{\text \bsle}(\tau)$
and is for given window function $f$ unique up to a constant phase.   
\end{theorem}

{\itshape Proof:} Let $S_{p}$ and $\tilde{S}_{p}$  be two fiducial solutions of \eqref{wave2}, related by a Bogoliubov transformation
\begin{eqnarray}\label{gsle5}
	S_{p}(\tau)\is a\tilde{S}_{p}(\tau)+b \tilde{S}_{p}(\tau)^\ast \,,\quad a,b\in \mathbb{C},\quad |a|^2-|b|^2=1\,.
\end{eqnarray}
In the following we shall omit the $p$-subscripts  for brevity. It is easy to to see that the functionals  $\mcE,\,\mcD$, as defined in \eqref{HHam3}, can be understood as the diagonal of two  bi-functionals, 
\begin{eqnarray}
	\mcE[S]\is C_1(S,S)\,,\quad \mcD[S]=C_2(S,S)\,.
\end{eqnarray}
Here $C_1$ is sesquilinear, $C_2$ is bilinear, and they are related by $C_1(S^\ast,S)=C_2(S,S)$ and $C_2(S^\ast,S)=C_1(S,S)$. Inserting the Bogoliubov transformation \eqref{gsle5}, application of  bi/sesqui-linearity yields
\begin{eqnarray}
	C_1(a \tilde{S}+b \tilde{S}^\ast,a \tilde{S}+b \tilde{S}^\ast) \is  |a|^2 C_1(\tilde{S},\tilde{S})+a^\ast b C_1(\tilde{S},\tilde{S}^\ast)+a b^\ast C_1(\tilde{S}^\ast, \tilde{S})+|b|^2 C_1(\tilde{S}^\ast,\tilde{S}^\ast)
	\nonum
	\is (|a|^2+|b|^2)C_1(\tilde{S},\tilde{S})+a^\ast b C_2(\tilde{S},\tilde{S})^\ast+a b^\ast C_2(\tilde{S}, \tilde{S})
	\nonum
	\is (|a|^2+|b|^2)\mcE[\tilde{S}]+a b^\ast \mcD[\tilde{S}]+a^\ast b\mcD[\tilde{S}]^\ast \,, 
	\\[2mm]
	C_2(a \tilde{S}+b \tilde{S}^\ast,a \tilde{S}+b \tilde{S}^\ast) \is  a^2 C_2(\tilde{S},\tilde{S})+a b C_2(\tilde{S},\tilde{S}^\ast)+a b C_2(\tilde{S}^\ast, \tilde{S})+b^2 C_2(\tilde{S}^\ast,\tilde{S}^\ast)
	\nonum
	\is a^2 C_2(\tilde{S},\tilde{S})+b^2 C_2(\tilde{S},\tilde{S})^\ast+2a b C_1(\tilde{S},\tilde{S})
	\nonum
	\is a^2 \mcD[\tilde{S}]+b^2\mcD[\tilde{S}]^\ast +2a b\mcE[\tilde{S}]\,.
\end{eqnarray}
Under a Bogoliubov transformation the functionals $\mcE,\,\mcD$ therefore transform as
\begin{eqnarray}\label{gsle7}
	\mcE[a \tilde{S}+b \tilde{S}^\ast]\is (|a|^2+|b|^2)\mcE[\tilde{S}]+a b^\ast \mcD[\tilde{S}]+a^\ast b\mcD[\tilde{S}]^\ast\,,
	\nonum
	\mcD[a \tilde{S}+b \tilde{S}^\ast]\is a^2 \mcD[\tilde{S}]+b^2\mcD[\tilde{S}]^\ast +2a b\mcE[\tilde{S}]\,.
\end{eqnarray}
In particular, 
\begin{eqnarray}\label{gsle8}
	\mcE[a \tilde{S}+b \tilde{S}^\ast]^2-|\mcD[a \tilde{S}+b \tilde{S}^\ast]|^2 \is
	\mcE[\tilde{S}]^2-|\mcD[\tilde{S}]|^2\,,\quad \quad \quad\,\,
\end{eqnarray}
i.e.~$\mcE[S]^2-|\mcD[S]|^2$ is a Bogoliubov invariant.

Next, consider the SLE  solution (\ref{sle5}) determined by the mode function $S_{p}$.
For its modulus-square one has 
 \begin{eqnarray}
	&&\big|T_{p}[S](\tau)\big|^2= (|\lambda_{p}[S]|^2+\mu_{p}[S]^2)|S_{p}(\tau)|^2
	+\lambda_{p}[S]\mu_{p}[S]S_{p}(\tau)^2 +\lambda_{p}[S]^\ast \mu_{p}[S]S_{p}(\tau)^{\ast\,2}
	\nonum
	&&\quad =\frac{1}{2\sqrt{\mcE_{p}[S]^2-|\mcD_{p}[S]|^2}}\big\{2\mcE_{p}[S]\big|S_{p}(\tau)\big|^2
	-\mcD_{p}[S]^\ast S_{p}(\tau)^2-\mcD_{p}[S] S_{p}(\tau)^{\ast \,2}\big\}\,.
\end{eqnarray}
Inserting the transformation \eqref{gsle5}, together with \eqref{gsle7}, \eqref{gsle8} yields
\begin{eqnarray}
	&& \big|T_{p}[a \tilde{S}+b \tilde{S}^\ast](\tau)\big|^2 
\nonum
	&& \quad = \frac{|a|^2-|b|^2}{2\sqrt{\mcE_{p}[\tilde{S}]^2-|\mcD_{p}[\tilde{S}]|^2}}\Big\{2\mcE_{p}[\tilde{S}]\big|\tilde{S}_{p}(\tau)\big|^2
	-\mcD_{p}[\tilde{S}]^\ast \tilde{S}_{p}(\tau)^2
	-\mcD_{p}[\tilde{S}] \tilde{S}_{p}(\tau)^{\ast \,2}\Big\}
	\nonum
	&& \quad = (|a|^2-|b|^2)\big|T_{p}[\tilde{S}](\tau)\big|^2
= \big|T_{p}[\tilde{S}](\tau)\big|^2\,,
\end{eqnarray}
i.e.~the SLE corresponding to fiducial mode functions $S$ and $\tilde{S}$ differ by at most by $\tau$ independent phase. This establishes the Bogoliubov invariance of the Wightman function \eqref{bsle3}. \qed

\subsection{\bsle in terms of commutator function} 
\label{subsec2c}

The ``commutator functional'' is another natural Bogoliubov invariant. Indeed,  
consider $\Delta[S](\tau,\tau_0) = i (S(\tau) S(\tau_0)^* - S(\tau)^* S(\tau_0) )$,
as a map $\Delta: C([\tau_i,\tau_f]) \rightarrow C([\tau_i,\tau_f]^2)$,
with $C([\tau_i,\tau_f])$ the continuous functions on $[\tau_i,\tau_f]$. 	
Then $\Delta[S](\tau,\tau_0)$ is real valued, antisymmetric in $\tau,\tau_0$, and obeys 
$\Delta[a S + b S^*](\tau,\tau_0) = (|a|^2 - |b|^2 ) \Delta[S](\tau,\tau_0)$, $a,b \in \bbC$. 
On a solution $S_p$ of (\ref{wave2}) $\Delta[S_p]$ becomes 
the commutator function $\Delta_p[\tau,\tau_0]$, which is independent of the choice of 
Wronskian normalized fiducial solution. The commutator function can also be characterized 
directly as the solution of 
\begin{eqnarray} 
\label{commutator} 
[ \pp_{\tau}^2 + \om_p(\tau)^2 ] \Delta_p(\tau,\tau_0) =0 \,, \quad 
\Delta_p(\tau,\tau_0) = - \Delta_p(\tau_0, \tau) \,, \quad
\pp_{\tau} \Delta_p(\tau,\tau_0) \big|_{\tau = \tau_0} = 1\,. 
\end{eqnarray}  
The solution of the initial value problem for (\ref{wave2}) can be   written in terms 
of the commutator function as  
\begin{equation}
\label{ssle1} 
T_p(\tau) = \Delta_p(\tau,\tau_0) \pp_{\tau_0} T_p(\tau_0) 
- \pp_{\tau_0} \Delta_p(\tau, \tau_0) T_p(\tau_0)\,.
\end{equation} 
In the following we seek to reinterpret the minimization procedure underlying the SLE 
as one with respect to the initial data in (\ref{ssle1}). If feasible, this will result in a 
(nonlinear) functional of the commutator function, thereby elucidating the Bogoliubov invariance 
of the \bsle.  

We begin by inserting (\ref{ssle1}) and its time derivative 
into the definitions of $\mcE_p, \,\mcD_p$ to obtain
\begin{eqnarray}
\label{ssle2} 
\mcE_p \is J_p(\tau_0) |w_p|^2 + K_p(\tau_0) |z_p|^2 
- \pp_{\tau_0} J_p(\tau_0) \Re(w_pz_p)\,, 
\nonum
\mcD_p \is J_p(\tau_0) w_p^2 + K_p(\tau_0) z_p^2 
- \pp_{\tau_0} J_p(\tau_0) w_pz_p \,,
\end{eqnarray} 
with $z_p:= T_p(\tau_0)$, $w_p := \pp_{\tau_0} T_p(\tau_0)$, 
subject to $w_p z_p^* - w_p^* z_p = -i$. The coefficients 
\begin{eqnarray} 
\label{ssle3}
J_p(\tau_0) \is \frac{1}{2} \int\! d\tau \, f(\tau)^2 
\big[ \big( \pp_{\tau} \Delta_p(\tau,\tau_0) \big)^2 + 
\om_p(\tau)^2 \Delta_p(\tau,\tau_0)^2 \big] \,,
\nonum
K_p(\tau_0) \is \frac{1}{2} \int\! d\tau \, f(\tau)^2 
\big[ \big( \pp_{\tau} \pp_{\tau_0} \Delta_p(\tau,\tau_0) \big)^2 + 
\om_p(\tau)^2 \big(\pp_{\tau_0} \Delta_p(\tau,\tau_0)\big)^2 \big]\,, 
\end{eqnarray} 
are manifestly positive and are Bogoliubov invariants in the above sense.  
No fiducial solution enters, instead $\mcE_p, \mcD_p$ in 
(\ref{ssle2}) are functions of the constrained complex initial data $z_p,w_p$.      
To proceed, we anticipate the inequality 
\begin{equation} 
\label{ssle4} 
4 K_p(\tau_0) J_p(\tau_0) - (\pp_{\tau_0} J_p(\tau_0))^2 >0\,.
\end{equation}
It is convenient to momentarily simplify the notation by writing 
$K,J,\dot{J}$ for $K_p(\tau_0), J_p(\tau_0)$, $\pp_{\tau_0} J_p(\tau_0)$,
respectively, and to also omit the subscripts $p$ from 
$z_p,w_p, \mcE_p, \mcD_p$. By multiplying $T_p(\tau)$ in (\ref{ssle1}) 
with a $\tau$-independent phase we may take
$z$ to be real and positive. The solution of the Wronskian condition 
then reads
\begin{equation} 
\label{ssle5} 
w = w_R - \frac{i}{2 z}\,, \quad w_R, \,z >0\,.
\end{equation}
Inserted into the above $\mcE$ one is lead to 
minimize
\begin{equation} 
\label{ssle6} 
\mcE = J\Big( w_R^2 + \frac{1}{4 z^2} \Big) + K z^2 - 
\dot{J} zw_R\,,
\end{equation}
which gives
\begin{equation} 
\label{ssle7}
(z^{\rm min})^2 = \frac{J}{\sqrt{ 4 KJ - \dot{J}^2}}\,,\quad 
w^{\rm min}_R = \frac{z^{\rm min}}{2} \frac{\dot{J}}{J}\,.
\end{equation}
Using 
\begin{equation} 
\label{ssle8} 
\frac{w^{\rm min}}{z^{\rm min}} 
= \frac{\dot{J}}{2 J} - i \frac{\sqrt{ 4 KJ - \dot{J}^2}}{2 J},
\end{equation}
one sees that the minimizer is also a zero of $\mcD$ in \eqref{ssle2}. The solution (\ref{ssle7}) is 
unique up to a constant phase fixed by choosing $z>0$. Reinserting (\ref{ssle7}) 
into $\mcE$ gives 
\begin{equation} 
\label{ssle9} 
\mcE^{\rm min} = \frac{1}{2} \sqrt{4 KJ - \dot{J}^2}\,. 
\end{equation}
Since $\mcE$ in the original form (\ref{HHam3}) is manifestly positive this verifies the selfconsistency 
of (\ref{ssle4}). Upon insertion of (\ref{ssle3}) in (\ref{ssle7}), (\ref{ssle8}) the minimizing initial data become 
functionals of $\Delta$, for which we write $z_p[\Delta](\tau_0) 
= z^{\rm min}$, $w_p[\Delta](\tau_0) = w^{\rm min}$. 

\begin{theorem} \label{sslethm} The \bsle solution can be expressed solely in terms 
	of the commutator function as follows:
		\begin{eqnarray}
		\label{ssle10}  
		T^{\text \bsle}_p(\tau) \is \Delta_p(\tau,\tau_0) w_p[\Delta](\tau_0) - 
		\pp_{\tau_0} \Delta_p(\tau, \tau_0) z_p[\Delta](\tau_0)\,,
		\nonum
		z_p[\Delta](\tau_0) \is \sqrt{\frac{J_p(\tau_0)}{ 2 \mcE_p^{\text \bsle}}} = 
		T_p^{\text \bsle}(\tau_0)\,,
		\nonum
		w_p[\Delta](\tau_0) \is \pp_{\tau_0} T_p^{\text \bsle}(\tau_0) - i 
		\sqrt{ \frac{\mcE_p^{\text \bsle}}{ 2 J_p(\tau_0)}} = (\pp_{\tau} T_p^{\text \bsle})(\tau_0)\,.
		\end{eqnarray}
		For the modulus and the phase this gives  
		\begin{equation}
		\label{ssle12}
		\big| T^{\text \bsle}_p(\tau)\big|^2 = \frac{J_p(\tau)}{2 \mcE_p^{\text \bsle}}\,, 
		\quad 
		\exp\{i\, {\rm arg}\, T_p^{\text \bsle}(\tau)\} = 
		\bigg(\frac{J_p(\tau,\tau_0) - i \mcE_p^{\text \bsle} \Delta_p(\tau,\tau_0)}{J_p(\tau,\tau_0) + i \mcE_p^{\text \bsle} \Delta_p(\tau,\tau_0) }\bigg)^{1/2}.
		\end{equation}
		Here $J_p(\tau_0) = J_p(\tau_0,\tau_0)$ with 
		\begin{equation} 
		\label{ssle11a}
		J_p(\tau,\tau_0) = \frac{1}{2} \int\! d\tau_1 \, f(\tau_1)^2 
		\big[\pp_{\tau_1} \Delta_p(\tau_1,\tau) \pp_{\tau_1} \Delta_p(\tau_1,\tau_0) + 
		\om_p(\tau_1)^2 \Delta_p(\tau_1,\tau) \Delta_p(\tau_1,\tau_0)\big] \,,
		\end{equation}
		and  
		$\mcE_p^{\text \bsle}$ is the minimal energy given by 
		\begin{eqnarray} 
		\label{ssle11}
		\big(\mcE_p^{\text \bsle}\big)^2 \is \frac{1}{8} \int d\tau d \tau' 
		f(\tau)^2 f(\tau')^2 \Big\{ 
		\big(\pp_{\tau} \pp_{\tau'} \Delta_p(\tau,\tau') \big)^2 + 
		2 \om_p(\tau')^2 (\pp_{\tau} \Delta_p(\tau',\tau) \big)^2 
		\nonum
		&+& \om_p(\tau)^2 \om_p(\tau')^2 \Delta_p(\tau,\tau')^2 \Big\}\,.
		\end{eqnarray} 
\end{theorem} 

Concerning the derivation, Eq.~(\ref{ssle10}) is the explicit form of (\ref{ssle1}) with 
minimizing parameters (\ref{ssle7}), (\ref{ssle9}). In the expressions (\ref{ssle11}), (\ref{ssle12}), 
or (\ref{ssle11a}), in a first step terms fourth or third order in $\Delta$ and its derivatives appear. 
Using differential identities for products of the commutator function these expressions can be 
simplified to the ones presented. The details follow those in \cite{BonusSLE}, proof of Theorem II.2b, and are omitted. 

By construction, the parameterizations  (\ref{ssle10}) and (\ref{sle5}) of the \bsle are 
equivalent. For a generic solution $T_p(\tau)$ one can match the 
parameterizations (\ref{sle1}) and (\ref{ssle1}) by realizing 
the commutator function in terms of $S$. This 
gives 
\begin{eqnarray} 
\label{ssle16} 
\lb = i \big(S_p(\tau_0)^* w - \pp_{\tau_0} S_p(\tau_0)^* z \big) \,,
\quad
\mu = i \big(\pp_{\tau_0} S_p(\tau_0) z - S_p(\tau_0) w \big) \,.
\end{eqnarray}   
One can show that -- modulo phase choices -- the minimizers
$\lb^{\rm min}, \mu^{\rm min}$ and $z^{\rm min}, w^{\rm min}$ are 
linked by the same relation. The phases are however not necessarily matched, 
in particular real $\mu$ does not automatically correspond to real $z$. 
\medskip


\section{The Hadamard Property}
\label{sec3}

Throughout Sections \ref{sec3}--\ref{sec5} we use a generic ``quadratic dispersion relation'' 
in the following sense. 
\begin{definition} \label{quaddispersion} 
	Let $J\subseteq \bbR$ be an open time interval, $\om_0: J \rightarrow \R$
	smooth, and $\om_2^{ij} = \om_2^{ji}: J \rightarrow \R$, for $i,j =1,\ldots, d$, smooth functions 
	such that $\om_2^{ij}(\tau) \wp_i \wp_j > 0$, for all $\wp \in \bbS^{d-1}$, $\tau\in J$.  
	Define $p:=\wp |p|\in \bbR^d$, for $\wp\in \bbS^{d-1}$ and $|p|\geq 0$, and set $\om_2(\tau, \wp)^2 
	:= \om_2^{ij}(\tau) \wp_i \wp_j$. Then  $\om_p(\tau),
	\tau \in J, \,p \in \R^d$ is called a quadratic dispersion if 
	\be 
\label{dispersion}
	\om_p(\tau) = \big[ \om_0(\tau)^2 + |p|^2 \om_2(\tau,\wp)^2\big]^{1/2}\,.
	\ee
\end{definition} 	

Since we take the spatial metric $\wg_{ij}$ to be smooth and positive definite, 
the Bianchi I dispersion $\om_p^{\mfb}$ in (\ref{Bdispersion}) is of course 
a special case. It is however useful to allow $\om_p \neq \om_p^{\mfb}$ for several 
reasons:

(i) Background field expansion: as noted after (\ref{action1}) an expansion 
around a time dependent scalar background field $\vp(t)$ leads to 
a time dependent mass-squared term $U''(\vp)$ which modifies $\om_p^{\mfb}$. 
Treating the geometry as nondynamical this no longer adheres to the 
minimal coupling principle.

(ii) Complications due to nonminimal curvature coupling: if the scalar field is 
non-minimally coupled to gravity a $-n \wg\, \xi R \phi^2$, $\xi \in \R$,  term is added to the expression in 
curly brackets in (\ref{action1}). For the Bianchi~I geometries (\ref{metric1}) the 
scalar curvature $R$ is a function of time only, and the wave equation (\ref{wave2}) 
has the mass term in (\ref{Bdispersion}) replaced with $\om_0(\tau)^2 = \wg(\tau) (m^2 + \xi R(\tau))$. As long as $\om_0(\tau)^2 \geq 0$ this is still a quadratic dispersion 
and the \bsle construction 
starting from (\ref{Bsle2}) will produce exact Hadamard states, as proven later in this section. 
On the other hand, for $\xi \neq 0$, the functional $E_p$ in (\ref{Bsle2}) no 
longer coincides with the (spatially Fourier transformed) time-time component $T_{00}$ of the energy-momentum tensor. The latter contains a first order derivative term of 
the squared field $\phi$, which results in a $\dd_{\tau} |S_p(\tau)|^2$ modification 
of $\cE_p[S]$ and a $\dd_{\tau} S_p(\tau)^2$ modification of $\cD_p[S]$. 
As a consequence, the constraint $\cE_p[S] > |\cD_p[S]|$ is nontrivial 
and restricts $\xi$ and/or the geometry. Subject to the   $\cE_p[S] > |\cD_p[S]|$
constraint the SLE minimization procedure carries over and leads to 
$\mu_p, \lb_p$ as in (\ref{sle3}). The Hadamard property based on the 
$T_{00}$ induced functionals $\cE_p[S], \cD_p[S]$ has for $\xi \neq 0$ not 
fully been proven, not even for \FL geometries.  Since in a non-minimally coupled
theory the energy-momentum tensor is not expected to satisfy the weak energy 
condition, one may also question the naturalness of the conditions enforcing 
positivity in all frames.

(iii) Inclusion of effective quantum gravity corrections: as mentioned 
in the Introduction effective loop quantum gravity effects can be attributed 
to a modified time dependent mass \cite{LQCSLE1,LQCSLE2}, motivating again to take $\om_p \neq \om_p^{\mfb}$.   

(iv) Use in the framework of the Functional Renormalization Group (FRG): strict 
1-loop Wilsonian renormalization of a scalar field theory on a non-static 
background \cite{FLFRG} requires in general the inclusion of an infinite tower 
of Ricci scalar terms, $U(\chi) \mapsto \sum_{n \geq 0} {}^nU(\chi) R^n$. The formally associated energy momentum tensor is complicated 
and of questionable physical significance, aggravating the problems in (ii).  
Since in most FRG applications composite operators are not needed, the 
use of the `minimal energy' functional in (\ref{Bsle2}) is warranted
and for a spatial mode modulation leads to generalized dispersion 
relations of the form   
\be 
\label{gendispersion}
\om_p(\tau)^2 = \wg(\tau) U''(\vp) + 
\wg(\tau) \Big[\wg^{ij}(\tau) p_i p_j
+ k^2 r\Big(\frac{\wg^{ij}(\tau) p_i p_j}{k^2}\Big) \Big]\,,
\ee
where $r \in \cS(\R_+)$ is a Schwartz function. Being of rapid decay, the 
$r$ term will not interfere with the asymptotic expansions in $1/|p|$ 
entering the computational framework of the flow equations, see \cite{FLFRG}.   
Although microlocal analysis applies to pseudo differential operators as well, 
and a Green's function analysis has begun \cite{GreensNonlocal}, 
an agreed upon generalization of the Hadamard property does not seem to be available. 
\medskip 

From now on we consider the basic wave equation (\ref{wave2}) with a generic 
quadratic dispersion relation. We note a simple consequence of the definition: for any compact time interval $I\subseteq J$,
there exist constants $0< c_- \leq c_+ < \infty$ such that
\be 
\label{om2bounds} 
c_- \leq \om_2^{ij}(\tau) \wp_i \wp_j \leq c_+\,, \quad \wp \in \bbS^{d-1}\,, \;\;
\tau \in I\,. 
\ee
Indeed, $\om_2(\tau)$ is pointwise in $\tau$ a symmetric, positive definite 
matrix. As such it can be diagonalised via an orthogonal matrix $O(\tau) \in O(d)$,
i.e.~$\om_2(\tau) = O(\tau)^T D(\tau) O(\tau)$, with $D(\tau) = 
(\lb_1(\tau),\ldots, \lb_d(\tau))$ collecting the positive eigenvalues. 
Thus $ \wp^T \om_2(\tau) \wp = \wp^O(\tau)^T D(\tau) \wp^O(\tau)$, with 
$\wp^O(\tau) := O(\tau)  \wp  \in \bbS^{d-1}$, can be bounded as 
\be 
\min_i \{ \lb_i(\tau) \}  \leq \wp^O(\tau)^T D(\tau) \wp^O(\tau) \leq 
\max_i \{ \lb_i(\tau) \} \,, 
\ee
using $\wp^O(\tau)^T \wp^O(\tau)=1$. Since $I$ is compact the infima and suprema
\be 
c_- = \inf_{\tau \in I}  \min_i \{ \lb_i(\tau) \} \,, \quad 
c_+ = \sup_{\tau \in I}  \max_i \{ \lb_i(\tau) \}\,,
\ee 
are assumed within the interval and thus obey $0 < c_- \leq c_+ < \infty$.
This gives (\ref{om2bounds}).  

As mentioned earlier, the selection of a homogeneous quasi-free state $\omega$ upon which to base perturbation theory is essentially 
equivalent to the choice of a solution to the Klein-Gordon equation; 
i.e.~in Bianchi I geometries (after spatial Fourier transform) to selecting 
a solution $S_p$ of 
$[ \partial_{\tau}^2 + \omega_p(\tau)^2 ] S_p(\tau) =0$, 
$(\partial_{\tau} S_p \,S_p^*)(\tau) - (\partial_{\tau} S_p^* \,S_p)(\tau) = -i$. 
The adiabatic vacua $\vacw{n}_p$ of Appendix \ref{secAdvac}  are such solutions, but generally lack the Hadamard property. 
In the following we establish a link between the \bsle and the adiabatic vacuum states, which will be essential 
for the proof that the \bsle are Hadamard states. By using adiabatic vacua $S_p^{(n)}$ of order $n\in \bbN_0$ as 
the fiducial solution for the construction of the \bsle we have the following lemma.
\begin{lemma} \label{lmbsad}
	Let $\om^\text{\bsle}$ be the State of Low Energy associated  to a smooth window function $f$ with compact support. Let $S_p^{(n)}(\tau)$ be the mode function associated to an adiabatic vacuum state of order $n\in \bbN_0$. 
	Then the mode function $T_p^{\text \bsle}(\tau)$ associated to $\om^\text{\bsle}$ may be realized as $T_p^{\text \bsle}(\tau)=\lambda_p[S^{(n)}] S_p^{(n)}(\tau)+
	\mu_p[S_p^{(n)}] S_p^{(n)}(\tau)^\ast$, i.e.~(\ref{sle5}) for $S_p = S_p^{(n)}$. 
	Further, uniformly over compact intervals of time,
	\begin{eqnarray}\label{bsad1}
	\pp_\tau^\ell T^{\text \bsle}_p(\tau)\is -\exx{-i \arg c^{(n)}_2}\pp_\tau^\ell S_p^{(n)}(\tau)+O(\mdp^{\ell-2n-1/2})\,,\quad \ell\in \bbN_0\,,
	\end{eqnarray}
	as $\mdp\to \infty$.	
\end{lemma}

{\itshape Proof:} Only (\ref{bsad1}) requires proof. For $S_p = S_p^{(n)}$ 
we write 
\begin{eqnarray}
c_1^{(n)}\is \frac{1}{2} \int\! d\tau f(\tau)^2 
\big[ |\partial_{\tau} S_p^{(n)}|^2 + \omega_p^2 |S_p^{(n)}|^2 \big]\,,
\nonum
c_2^{(n)}\is \frac{1}{2} \int\! d\tau f(\tau)^2 
\big[ (\partial_{\tau} S_p^{(n)})^2 + \omega_p^2 (S_p^{(n)})^2 \big]\,.
\end{eqnarray}
Since $c^{(n)}_1>|c^{(n)}_2|$, we have have from \eqref{sle3}, \eqref{sle4} the simple bound
\begin{eqnarray}\label{bsad1aa}
\mu_p^2\leq \frac{1}{2}\frac{|c^{(n)}_2|^2/(c^{(n)}_1)^2}{1-|c^{(n)}_2|^2/(c^{(n)}_1)^2}\,.
\end{eqnarray}
For an adiabatic vacuum state $S_p^{(n)}(\tau)$ as the fiducial solution, 
it follows from Theorem \ref{thad1} that there is a constant $A_1>0$ such that 
\begin{eqnarray}
c_1^{(n)}\geq A_1 \int\! d\tau f(\tau)^2 
\big[ |\partial_{\tau} W_p^{(n)}|^2 + \omega_p^2 |W_p^{(n)}|^2 \big]\,.
\end{eqnarray}
Then the upper and lower bounds \eqref{iter01d}, \eqref{iter01e} on $\Omega_p^{(n)}$, together with the definition \eqref{advac1} entails  there is a constant $A>0$ such that 
\begin{eqnarray}\label{bsad1a}
c^{(n)}_1\geq A(1+\mdp)\,,\quad \mdp\to \infty\,.
\end{eqnarray}
Next $c_2^{(n)}$ may be rewritten via Theorem \ref{thad1} as
\begin{eqnarray}\label{bsad2}
c_2^{(n)}\is \frac{1}{2} \int\! d\tau f(\tau)^2 
\big[ (\partial_{\tau} W_p^{(n)})^2 + \omega_p^2 (W_p^{(n)})^2 \big]+O(\mdp^{-2n+1})\,,
\end{eqnarray}
where explicitly
\begin{equation}
(\partial_{\tau} W_p^{(n)})^2 + \omega_p^2 (W_p^{(n)})^2 = \bigg[\Big(\frac{\pp_\tau \Omega_p^{(n)}}{2\Omega_p^{(n)}}+i\Omega_p^{(n)}\Big)^2+\frac{\omega_p^2}{2\Omega_p^{(n)}}\bigg]\frac{1}{2 \Omega_p^{(n)}}\exx{-2i\int_{\tau_0}^\tau\Omega_p^{(n)}(s)ds}\,.\quad 
\end{equation}
By performing repeated integrations by parts in \eqref{bsad2} through the identity
\begin{eqnarray}
\exx{-2i\int_{\tau_0}^\tau\Omega_p^{(n)}(s)ds}\is \frac{i}{2\Omega_p^{(n)}(\tau)}\pp_\tau\Big(\exx{-2i\int_{\tau_0}^\tau\Omega_p^{(n)}(s)ds}\Big)\,,
\end{eqnarray}
one sees that $\int\! d\tau f(\tau)^2 
[ (\partial_{\tau} W_p^{(n)})^2 + \omega_p^2 (W_p^{(n)})^2 ]$ is of rapid descent as $\mdp\to \infty$. Hence \eqref{bsad2} reduces to $c^{(n)}_2 = O(\mdp^{-2n+1})$. In combination with \eqref{bsad1a} this gives 
\begin{eqnarray}\label{bsad3}
|c_2^{(n)}|/c_1^{(n)}=O(\mdp^{-2n})\,.
\end{eqnarray}
Together, we have $\mu_p=O(\mdp^{-2n})$ and $|\lb_p|=(1+\mu_p^2)^{1/2}=1+O(\mdp^{-2n})$. The result \eqref{bsad1} then follows from \eqref{sle1} in combination with the bounds of Corollary \ref{corad1}.
\qed

\medskip

Now we claim that the \bsle have the Hadamard property:
 \begin{theorem} \label{thHad}
 	Let $\om^\text{\bsle}$ be the homogeneous pure quasifree state associated to the \bsle mode functions 
 	$T^{\text \bsle}_{p}$ in Theorem \ref{thbogo}. Then $\om^\text{\bsle}$ has the Hadamard property.
 \end{theorem}
 Radzikowski \cite{HadamardRadzikowski} gave a geometrical reformulation of the  original definition of a Hadamard state by Kay and Wald \cite{HadamardWald1} in terms  of the wavefront set of its two-point distribution.
 \begin{definition} \label{DefHadWF}
Consider the Klein-Gordon equation on a globally hyperbolic manifold $M$, with algebra of observables $\mathcal{A}(M)$. A quasifree state $\omega$ on the $\mathcal{A}(M)$ is said to be a {\itshape Hadamard state} if the wavefront set of its two-point distribution $\varpi$ satisfies $\text{WF}(\varpi)=C^+$, where $C^+$ is a subset of $T^\ast (M\times M)$, the cotangent bundle over $M\times M$, defined as 
	\begin{eqnarray}\label{WFhad}
		C^+:= \big\{(y_1,\xi_1;y_2,-\xi_2)\in T^\ast (M\times M)\setminus\{0\}\big\vert (y_1,\xi_1)\sim (y_2,\xi_2),\,\xi_1\in \overline{V}_+\big\}\,.
	\end{eqnarray}
	Here the relation $(y_1,\xi_1)\sim (y_2,\xi_2)$ means that there exists a null-geodesic in $M$ connecting $y_1$ and $y_2$, $\xi_1$ is the covector to this geodesic at $y_1$, and $\xi_2\in T_{y_2}^\ast M$ is the parallel transport of $\xi_1$ along the geodesic. Finally, $\overline{V}_+$ is the closed forward lightcone of $T_{y_1}^\ast M$.
\end{definition}	 
The wavefront set of a distribution captures not only the location of the distribution's singularities, but also the direction(s) in which they are propagated. Although formulated more abstractly than the original definition in terms of the Hadamard parametrix, this microlocal definition is more suitable for mathematical analysis.
Our proof that the \bsle are Hadamard states, i.e. $\text{WF}(\varpi^\text{\bsle})=C^+$, proceeds along by-now standard lines \cite{Olbermann, SLE2, SJHadamard}. In order to show that the wavefront set of the \bsle coincides with that of a generic Hadamard state, we use  adiabatic vacuum states as a ``conduit'' by appealing to a result of   Junker-Schrohe \cite{JunkerS} that establishes a connection between the Sobolev wavefront sets of adiabatic vacuum  states and a generic Hadamard state. Specifically, we shall show that the Sobolev wavefront sets of the \bsle coincide with that of an adiabatic vacuum of sufficiently high order, and hence (by the aforementioned result) coincide with the Sobolev wavefront sets of a generic Hadamard state.

\subsection{A brief review of wavefront sets}
\label{subsec3a}

For convenient reference we collect some salient aspects of wavefront sets required for the proof, referring to \cite{KhavMoretti,Gerardbook,microref} for further details.
In order to introduce the notion of the wavefront set for distributions on manifolds, it is instructive to first consider a distribution $u\in \mcD'(\bbR^D)$.\footnote{Here $\mathcal{D}'(\mathbb{R}^D)$, the space of distributions, is the topological dual of $\mathcal{D}(\mathbb{R}^D)\equiv C_c^\infty(\mathbb{R}^D)$.} Information about the location of the singularities of $u$, together with the directions in which they propagate, is contained in the wavefront set $\text{WF}(u)$. To define this, the distribution is first localized about some $y_0\in \bbR^D$ via a $h\in C_c^\infty(\bbR^D)$ with $h(y_0)\neq 0$, yielding a distribution $hu$ of compact support. Next, one defines the set of {\itshape regular directed points} of $u$ as those points $(y_0,\xi_0)\in \bbR^D\times \bbR^D\setminus\{0\}$ for which there exists $h\in C_c^\infty(\bbR^D)$ and an open conic neighborhood $\Gamma_{\xi_0}$ of $\xi_0\in \bbR^D\setminus\{0\}$ such that for each $n\in \bbN$ there exists a constant $C_n>0$ with 
\begin{eqnarray}\label{gslehad1}
	\sup_{\xi\in \Gamma_{\xi_0}}\big|(hu)^\wedge(\xi)\big|\leq C_n (1+|\xi|)^{-n}\,.
\end{eqnarray}
Here $(hu)^\wedge$ is the Fourier transform of $hu\in \mcD'(\bbR^D)$ and $|\cdot|$ is the  usual Euclidean norm  on $\bbR^D$. 
Then the $C^\infty$-{\itshape wavefront set} $\text{WF}(u)$ is defined as the complement of the set of regular directed points in  $\bbR^D\times \bbR^D\setminus\{0\}$. This set is closed, and transforms as a subset of the cotangent bundle under coordinate transformations \cite{Hormander1}. The local character of this construction allows a straightforward generalization of wavefront sets to distributions $u$ on a smooth manifold $N$  using charts. Moreover, since in local coordinates $\text{WF}(u)$ transforms as a subset of $T^\ast N$ under coordinate changes, WF$(u)$ is independent of the particulars of the charts chosen.   

Finally, we note an  important property of wavefront sets concerning sums of distributions, namely
\begin{eqnarray}\label{gslehad2}
	\text{WF}(u+v)\subseteq \text{WF}(u)\cup \text{WF}(v)\,.
\end{eqnarray}
While the Hadamard property is conveniently encapsulated within the  $C^\infty$-wavefront set, the description of adiabatic vacuum states requires the introduction of a further refinement, the Sobolev wavefront sets $\text{WF}^s$. In addition to a singularity's location and direction of propagation, they contain information about the {\itshape degree} of the singular behavior. Junker and Schrohe \cite{JunkerS} utilized Sobolev wavefront sets to   rigorously define adiabatic vacuum states on general globally hyperbolic manifolds, as well as characterize their relation to Hadamard states. An essential notion for the definition of the Sobolev wavefront set is that of  a {\itshape local Sobolev space} $H^s_\text{loc}$. 

\begin{definition}[Local Sobolev spaces] 
Let $N$ be a $D$-dimensional smooth manifold. We say that $u\in \mathcal{D}'(N)$ is in $H^s_\text{loc}(N)$ for $s\in \mathbb{R}$ if for any chart $(U,\Phi)$ and any $h\in C_c^\infty(U)$
	\begin{eqnarray}\label{sob3}
	\int\!d^D\xi\,(1+|\xi|^2)^s |(\Phi^\ast (h u))^\wedge(\xi)|<\infty\,,
\end{eqnarray}
with $\Phi^\ast$ denoting the pull-back. Recall that $\mcD'(N)$ is the topological dual of $C_c^\infty(N)$,  $(\Phi^\ast (h u))^\wedge$ is the Fourier transform of the pull-back distribution $\Phi^\ast (h u)\in \mcD'(\bbR^D)$, and $|\cdot|$ is the usual Euclidean norm on $\bbR^D$.

\end{definition}

The Sobolev wavefront sets on $\bbR^D$ are defined as usual through a complement:

\begin{definition}[Sobolev wavefront sets on $\bbR^D$]\ 
Let $u\in \mathcal{D}'(\mathbb{R}^D)$, $y_0\in \mathbb{R}^D,\,\xi \in \bbR^D\setminus\set{0}$ and $s\in \bbR$. Then $(y_0,\xi_0)\notin \text{WF}^s(u)$ if there exists $h\in C_c^\infty(\bbR^D)$ with $h(y_0)\neq 0$ and an open conic neighborhood $\Gamma$ of $\xi_0\in \bbR^D\setminus\set{0}$  such that 
	\begin{eqnarray}
	\int_\Gamma \!d^D\xi \,(1+|\xi|^2)^s |(\vp u)^\wedge(\xi)|^2 <\infty\,.
\end{eqnarray}

\end{definition}
It was shown in \cite{JunkerS} that for distributions on any subset $O$ of $\bbR^D$, the corresponding Sobolev wavefront sets are subsets of the cotangent bundle $T^\ast O\setminus\{0\}$, and further that this definition can be extended to general (paracompact) smooth manifolds using partitions of unity.

\noindent
{\bfseries Remarks.}

(i) It is clear that $\text{WF}^s(u)=\emptyset$ is equivalent to $u\in H^s_\text{loc}(N)$. It also follows from the expression \eqref{sob3} in the definition   of  local Sobolev spaces, as well as  standard properties of the Fourier transform on $\bbR^D$ that all  functions of sufficiently high differentiability are contained therein, specifically
	\begin{eqnarray}\label{sob4}
	\forall\,s<j-\half \text{dim}(N):\,C^j(N)\subseteq H^s_\text{loc}(N)\,,
\end{eqnarray}
with the $\frac{1}{2} \text{dim}(N)$ arising from the weight factor  in \eqref{sob3}.

(ii) The analog of \eqref{gslehad2} for the Sobolev wavefront sets is
\begin{eqnarray}\label{sob5}
	\forall\,s\in \bbR:\, \text{WF}^s(u+v)\subseteq \text{WF}^s(u)\cup \text{WF}^s(v)\,.
\end{eqnarray}

(iii) The $C^\infty$-wavefront set is related to the Sobolev wavefront sets by
  \begin{eqnarray}\label{sob6}
	\text{WF}(u)=\overline{\bigcup_{s\in \bbR}\text{WF}^s(u)}\,.
\end{eqnarray}
Finally, the relationship between adiabatic vacuum states and Hadamard states is characterized in terms of   Sobolev wavefront sets from \cite{JunkerS}:
\begin{lemma}[Lemma 3.3 from \cite{JunkerS}]\ \label{JSlemma}
Let $\varpi^{\rm H}$ and $\varpi^{(n)}$, respectively,  be the two-point functions of an arbitrary Hadamard state and an adiabatic vacuum of order $n\in \bbN_0$ associated to the Klein-Gordon field on a four dimensional globally hyperbolic manifold $(M,g)$. Then 
	\begin{eqnarray}
	\forall\,s<2n+\frac{3}{2}:\,\text{WF}^s\big(\varpi^{\rm H}-\varpi^{(n)} \big)=\emptyset\,.
\end{eqnarray}
\end{lemma}


\subsection{Proof of the Hadamard Property}
\label{subsec3b}

We now to turn to proving that the \bsle construction defines Hadamard states.  Suppose that $\varpi^{\rm H}$ is the two-point function of some Hadamard state. By \eqref{gslehad2}
 \begin{eqnarray}
	\text{WF}(\varpi^\text{\bsle})\is \text{WF}(\varpi^\text{\bsle}\!-\varpi^{\rm H}+ \varpi^{\rm H} )
	\subseteq \text{WF}(\varpi^\text{\bsle}\!-\varpi^{\rm H})\cup \text{WF}(\varpi^{\rm H})
	\nonum
	\is \text{WF}(\varpi^\text{\bsle}\!-\varpi^{\rm H})\cup C^+\,,
\end{eqnarray}
with $C^+$ defined in \eqref{WFhad}. Thus, to prove $\omega^\text{\bsle}$ is a Hadamard state, it is sufficient to show that $\text{WF}(\varpi^\text{\bsle}-\varpi^{\rm H})=\emptyset$. 
Specializing henceforth to $D=4$, fixing arbitrary $s\in \bbR$ and $n\in \bbN_0$ we have by \eqref{gslehad2}
\begin{eqnarray}
	\text{WF}^s(\varpi^\text{\bsle}\!-\varpi^{\rm H})\subseteq \text{WF}^s(\varpi^\text{\bsle}\!-\varpi^{(n)})\cup \text{WF}^s(\varpi^{(n)}-\varpi^{\rm H})\,,
\end{eqnarray} 
 and hence by Lemma \ref{JSlemma} 
\begin{eqnarray}\label{had2}
	\forall\,s<2n+\frac{3}{2}:\,\text{WF}^s(\varpi^\text{\bsle}\!-\varpi^{\rm H})\subseteq \text{WF}^s(\varpi^\text{\bsle}\!-\varpi^{(n)})\,.
\end{eqnarray}
 We shall now show that for every $s\in \mathbb{R}$ it is possible to choose $n\in\bbN_0$ sufficiently large so that the RHS of \eqref{had2} is empty. This together with \eqref{sob6} above establishes that $\text{WF}(\varpi^\text{\bsle}\!
 -\varpi^{\rm H})=\emptyset$.

Consider the Fourier realization of $\Delta \varpi^{(n)} := \varpi^\text{\bsle}\!-\varpi^{(n)}$,
\begin{eqnarray}\label{gslehad3}
	\Delta \varpi^{(n)}(\tau,x;\tau',x'):=\int\! \frac{d^3p}{(2\pi)^3}\,e^{ip(x-x')}[T_{p}^{\mfb\text{SLE}}(\tau)T_{p}^{\mfb\text{SLE}}(\tau')^\ast -S^{(n)}_{p}(\tau)S^{(n)}_{p}(\tau')^\ast ]\,.
\end{eqnarray}
We operate on it with arbitrary mixed spacetime derivatives  $\pp_{(y,y')}^\al \Delta \varpi^{(n)}(y,y')$, for some  
multi-index $\alpha$. Collecting temporal and spatial derivatives this gives a differential operator of the form  
$\dd_{\tau}^{a_1} \dd_{\tau'}^{a_2} \dd_{(x,x')}^{\alpha_s}$, with $|\alpha| = 
a_1 + a_2 + |\alpha_s|$. Momentarily averaging (\ref{gslehad3}) with test functions 
the repeated differentiations can be pulled inside the momentum integral. Validity for arbitrary test 
functions allows one to return to the pointwise version as long as the integrals encountered are 
absolutely convergent. Writing $\mcK_p^{(n)}(\tau,\tau'):= 
T_{p}^{\mfb\text{SLE}}(\tau)T_{p}^{\mfb\text{SLE}}(\tau')^\ast - S^{(n)}_{p}(\tau)S^{(n)}_{p}(\tau')^\ast$, leads to integrals of the form  
\begin{eqnarray}
\label{bslehad2}
\int\! \frac{d^3p}{(2\pi)^3}\, \pp_{(x,x')}^{\alpha_s}  e^{ip(x-x')} \pp_{\tau}^{a_1} \pp_{\tau'}^{a_2} 
\mcK_p^{(n)}(\tau,\tau')\,.
\end{eqnarray}
Now, it follows from Lemma \ref{lmbsad} that for any $a_1,a_2 \in \bbN_0$ we have 
\begin{eqnarray}\label{bslehad1}
	\pp_{\tau}^{a_1} \pp_{\tau'}^{a_2} \mcK_p^{(n)}(\tau,\tau')=O(\mdp^{a_1 + a_2-2n-1})\,,
\end{eqnarray}
uniformly over compact time intervals. Clearly, $\pp_{(x,x')}^{\alpha_s}  \exx{ip(x-x')}=O(\mdp^{|\alpha_s|})$. 
This gives the estimate 
\be
	\bigg|\int\! \frac{d^3p}{(2\pi)^3}\, \pp_{(x,x')}^{\alpha_s}  e^{ip(x-x')} \pp_{\tau}^{a_1} \pp_{\tau'}^{a_2} 
	\mcK_p^{(n)}(\tau,\tau') \bigg|
	\leq \text{const.}\int_0^\infty\!\!d|p|\,  (1+\mdp)^{|\al|-2n+1}\,,
\ee 
and thus the integral is absolutely convergent as long as $|\al|\leq  j(n):=2n-3$. 
Further, since the action of the derivative $\pp_{(y,y')}^\al$ under the integral in \eqref{gslehad3} is integrable for $|\al|\leq  j(n)$, it follows that $\Delta \varpi^{(n)}\in C^{j(n)}(M\times M)$, where $M$ here denotes the Bianchi I spacetime. By \eqref{sob4} this shows that $\forall\,s\in \bbR$, $\forall\,n\in \bbN_0$, if $s<j(n)-4=2n-7$ one has $\Delta \varpi^{(n)}\in H^s_{\text{loc}}(M\times M)$.
Equivalently, for every $s\in \mathbb{R}$ there exists an adiabatic order $n\in \mathbb{N}_0$ such that  $\text{WF}^s(\varpi^\text{\bsle}-\varpi^{(n)})=\emptyset$. Combined with \eqref{had2} this entails $\text{WF}^s(\varpi^\text{\bsle}\!-\varpi^{\rm H})=\emptyset$ for every $s\in \bbR$. 
Then, by  \eqref{sob6}
\begin{eqnarray}
	\text{WF}(\varpi^\text{\bsle}\!-\varpi^{\rm H})=\overline{\bigcup_{s\in \bbR}\text{WF}^s(\varpi^\text{\bsle}
		-\varpi^{\rm H})}=\emptyset\,,
\end{eqnarray}
which completes the proof that  $\omega^\text{\bsle}$ is a Hadamard state.
\qed


\section{Infrared behavior of \bsle states} 
\label{sec4}

While the ultraviolet behavior of all Hadamard states for a QFT 
is the same  (in principle solely determined by the Hessian) the 
infrared behavior is specific to the Hadamard state under consideration. 
As a consequence, information about the infrared behavior of 
a Hadamard state is only available in cases where it is obtained via a 
concrete construction principle. For the SLEs the minimization principle 
fixes the spatial momentum dependence completely, for {\it all} nonzero momenta,
including small. In this section we derive a {\it convergent} series expansion 
for the \bsle for small spatial momenta, applicable for any quadratic dispersion  
relation. It remains valid in the massless case 
and then produces a universal leading long distance decay 
\begin{eqnarray}
\label{sleIR} 
\int\! \!\frac{d^dp}{(2\pi)^d} \,T_p^{\mfb\text{SLE}}(\tau) T_p^{\mfb\text{SLE}}(\tau')^* 
e^{- i p(x-x')} = \frac{\Gamma( \frac{d-1}{2})}{4\pi^{\frac{d+1}{2}}} 
\frac{\det Q}{[(x-x')^i Q^{-1}_{ij} (x-x')^j]^{\frac{d-1}{2}}} + \ldots
\end{eqnarray}
where the dots indicate terms decaying faster in $|x-x'|$ and $Q^{ij}$ is 
(for the Bianchi I dispersion proper) a matrix determined solely by the 
spatial metric and the window function,
see (\ref{massless3}). For $Q= \delta^{ij}$ the displayed term is precisely 
the massless two-point function in Minkowski space, so {\it all} 
\bsle states qualitatively have the same, anisotropy modulated, Minkowski-like 
long distance decay.

\subsection{Solutions with strictly homogeneous initial data}\label{sec3.1} 

Since \bsle are independent of the choice of fiducial solution their expansion can be 
based on a conveniently chosen class of reference solutions.  We thus 
return to the basic wave equation (\ref{wave2}) and seek solutions $S_p(\tau)$ 
with infrared finite (strictly homogeneous) initial data, $\lim_{p_i \rightarrow 0} S_p(\tau_0) = z_0$,  $\lim_{p_i \rightarrow 0}\pp_{\tau_0}S_p(\tau_0) = w_0$. 
One might then try to find solutions via a multivariable Taylor series ansatz 
$\sum_{n \geq 0} a^{i_1\ldots i_n}(\tau) p_{i_1} \ldots p_{i_n}$.
While this is feasible, it turns out to be  more economical to define bounded `directional momenta' $\wp_i$ and an `overall momentum scale' $|p|$ as in Definition 
\ref{quaddispersion}, i.e.~$p_i := |p| \wp_i$, $|p| = \sqrt{\delta^{ij} p_i p_j}$.
For a generic quadratic dispersion we consider 
\begin{equation}
\label{odedef} 
[\partial_{\tau}^2 + \omega_p(\tau)^2] S_p(\tau) =0\,, \quad 
\omega_p(\tau)^2 = \omega_0(\tau)^2 + |p|^2 \omega_2(\tau,\wp)^2\,,  
\end{equation}
where $\omega_0(\,\cdot\,)$ is allowed to vanish while $\omega_2(\,\cdot\,,\wp)$ is 
bounded away from zero, see (\ref{om2bounds}). 
The following result ensures the existence of convergent series solutions of (\ref{odedef})
with strictly homogeneous  initial data.

\begin{proposition} \label{Ssmallp} The differential equation 
	(\ref{odedef}) admits convergent series solutions 
	with a radius of convergence $|p|_*>0$ on $[\tau_i,\tau_f]$, 
	such that for any $|p|<|p|_*$ 
	\begin{eqnarray}
	\label{Sseries} 
	S_p(\tau)&\! =\! & \sum_{n=0}^\infty S_n(\tau,\wp) |p|^{2n}\,,
	\quad {\rm and} \quad \partial_\tau S_p(\tau)=
	\sum_{n=0}^\infty \partial_\tau S_n(\tau,\wp)|p|^{2n}\,,
	\end{eqnarray}
	and the sums converge {\it uniformly} on $[\tau_i,\tau_f]$.   These solutions have IR finite initial data
	\begin{eqnarray}
	\label{SSeriesinit} 
	\lim_{p_i \rightarrow 0} S_p(\tau_0) =: z_0 < \infty\,,\quad 
	\lim_{p_i \rightarrow 0} \partial_{\tau_0} S_p(\tau_0) =: w_0 < \infty\,,
	\end{eqnarray}
and are fully determined by these data. As a consequence, the commutator function $\Delta_p(\tau,\tau')$ 
and the Green's functions defined in terms of it 
have uniformly convergent series expansions in $|p|< |p|_*$ as well, for 
distinct $(\tau, \tau') \in [\tau_i,\tau_f]\times [\tau_i,\tau_f]$. 
\end{proposition}

The proof is a straightforward generalization of Proposition~III.1 and Corollary~III.3 in \cite{BonusSLE} and is omitted. 

So far these are existence results. For the construction of 
these series solutions one will solve the recursion relations implied by the ansatz
(\ref{Sseries}) 
\begin{eqnarray}
\label{Srec1} 
[\partial_{\tau}^2 + \omega_0(\tau)^2 ]S_0(\tau) &\! =\! & 0\,, 
\nonumber \\[1.5mm]
[\partial_{\tau}^2 + \omega_0(\tau)^2 ]S_n(\tau,\wp) &\! =\! & - \omega_2(\tau,\wp)^2 S_{n-1}(\tau,\wp) \,, 
\quad n \geq 1\,.
\end{eqnarray}
In itself these determine some $S_n$, $n \geq 1$, only up to addition of a solution of the homogeneous 
equation, characterized by two complex parameters. The initial conditions 
(\ref{SSeriesinit}) eliminate this ambiguity as $S_n(\tau_0,\wp) =0$, $\pp_{\tau_0} S_n(\tau_0, \wp) =0$,
$n \geq 1$. This suggests to use $\tau_0 = \tau_i$, in which case the relevant Green's function is the 
retarded one, $G^{\wedge}_0(\tau,\tau') := \theta(\tau-\tau') 
\Delta_0(\tau,\tau')$, with $\Delta_0$ the commutator function for $\partial_{\tau}^2 + \omega_0(\tau)^2$. 
The solution of the iteration is then simply 
\begin{eqnarray}
\label{Srec3} 
S_n(\tau,\wp) \!\!&\! =\! & \!\!\int_{\tau_i}^{\tau_f} \! d\tau' \, 
K_n(\tau,\tau',\wp)\, S_0(\tau')\,, \quad n \geq 1\,,
\\[2mm] 
K_1(\tau,\tau',\wp)  \!&\!:=\!&\! - G_0^{\wedge}(\tau,\tau') \omega_2(\tau',\wp)^2\,,
\nonumber \\[1.5mm]
K_{n+1}(\tau,\tau') \!&\!:=\!&\! (-)^{n+1} 
\int_{\tau_i}^{\tau_f} \!\! d\tau_1...d\tau_n 
\, G_0^{\wedge}(\tau,\tau_1) \omega_2(\tau_1,\wp)^2 
\nonumber
\\[2mm]
\!&\!\times\!& \!G_0^{\wedge}(\tau_1,\tau_2) \omega_2(\tau_2,\wp)^2  \ldots 
\, G_0^{\wedge}(\tau_n,\tau') \omega_2(\tau',\wp)^2\,. 
\nonumber
\end{eqnarray}    
The kernel $K_n$ is manifestly real and satisfies $K_n(\tau_i, \tau') =0= 
\partial_{\tau} K_n(\tau,\tau')|_{\tau = \tau_i}$, for $\tau' \in (\tau_i, \tau_f]$. 
The associated series solution $S_p(\tau)$ therefore satisfies
$S_p(\tau_i) = z_0$, $(\partial_{\tau} S_p)(\tau_i) = w_0$, for $p$-independent 
constants with $w_0 z_0^* - w_0^* z_0 = -i$.    

A first application of (\ref{Srec3}) is to the construction of the 
commutator function. Recall from (\ref{commutator}) that $\Delta_p(\tau,\tau')$ is independent 
of the choice of the Wronskian normalized solution used to realize it. We are thus free 
to use the solution (\ref{Srec3}) for this purpose. Writing $\Delta_p(\tau,\tau') = 
\sum_{n\geq 0} \Delta_n(\tau,\tau',\wp) |p|^{2n}$, one finds 
\begin{eqnarray}
\label{Deltaexp1} 
\Delta_n(\tau,\tau',\wp) &\! =\! & i \sum_{j=0}^n \big(S_j(\tau,\wp) S^*_{n-j}(\tau',\wp) - 
S_j^*(\tau,\wp) S_{n-j}(\tau',\wp) \big) 
\nonumber \\[1.5mm]
&\! =\! & \int_{\tau_i}^{\tau_f} \! ds [ K_n(\tau,s,\wp)\Delta_0(s,\tau')  - 
K_n(\tau',s,\wp) \Delta_0(s,\tau)]
\nonumber \\[1.5mm]
&+& 
\int_{\tau_i}^{\tau_f} \! ds_1 ds_2 \, 
\sum_{j=1}^{n-1} K_j(\tau, s_1,\wp) K_{n-j}(\tau',s_2,\wp) \Delta_0(s_1,s_2) \,.
\end{eqnarray}
As a check one may verify that these coefficients satisfy the relations implied
by the expansion of the defining conditions (\ref{commutator})
\begin{eqnarray}
\label{Deltaexp2}
[\partial_{\tau}^2 + \omega_0(\tau)^2 ]\Delta_n(\tau,\tau',\wp) &\! =\! & - \omega_2(\tau)^2
\Delta_{n-1}(\tau,\tau',\wp) \,,
\quad
\partial_{\tau} \Delta_n(\tau, \tau',\wp) \big|_{\tau = \tau'} =0\,,
\nonumber \\[1.5mm]
[\partial_{\tau'}^2 + \omega_0(\tau')^2]\Delta_n(\tau,\tau',\wp) &\! =\! &
- \omega_2(\tau')^2 \Delta_{n-1}(\tau,\tau',\wp) \,, \;\quad n \geq 1\,.
\end{eqnarray}


\subsection{IR expansion of \bsle}\label{sec3.2}

We use the formulas from Theorem \ref{sslethm} to derive 
convergent series expansions for the \bsle. The basic expansion is that of 
$\Delta_p$ which entails analogous ones for $J_p$ and ${\cal E}_p^{\text{\bsle}}$: 
\begin{eqnarray}
\label{sleexp0}
\Delta_p(\tau',\tau) \is \sum_{n\geq 0} \Delta_n(\tau',\tau,\wp) |p|^{2n}\,,
\nonum
J_p(\tau',\tau) \is \sum_{n\geq 0} J_n(\tau',\tau,\wp) |p|^{2n}\,,
\quad 
({\cal E}_p^{\text{\bsle}})^2 = \sum_{n \geq 0} \varepsilon_n^2(\wp) |p|^{2n}\,. 
\end{eqnarray}
The uniform convergence of the various pointwise products is ensured by Lemma III.2 of 
\cite{BonusSLE} and allows one to exchange the order of summation and integration. 
For the `square' of a series $C_p(\tau,\tau_0) = \sum_{n \geq 0} C_n(\tau,\tau_0) \,|p|^{2n}$ a convenient notation is  
\begin{equation}
\label{sleexp1} 
C_p(\tau_1,\tau) C_p(\tau_1,\tau_0) = \sum_{n \geq 0} C(\tau_1|\tau,\tau_0)^2_n \,|p|^{2n} \,,
\quad C(\tau_1|\tau,\tau_0)^2_n := \sum_{j=0}^n 
C_j(\tau_1,\tau) C_{n-j}(\tau_1, \tau_0) \,.
\end{equation}
Applied to derivatives of the $\Delta_p$ series we suppress the residual $\wp_i$ 
dependence of the coefficients. In this notation one has 
\begin{eqnarray}
\label{sleexp2}
J_n(\tau,\tau_0,\wp) \is \frac{1}{2} \int\! d\tau_1 \, f(\tau_1)^2 
\Big\{( \partial_{\tau_1} \Delta)(\tau_1|\tau,\tau_0)^2_n + 
\omega_0(\tau_1)^2 \Delta(\tau_1|\tau,\tau_0)^2_n 
\nonum
&+& 
\omega_2(\tau_1,\wp)^2 \Delta(\tau_1|\tau,\tau_0)^2_{n-1} \Big\} \,,
\\[1.5mm]
\varepsilon_n^2(\wp) \is \frac{1}{8} \int\! d\tau d\tau' f(\tau^2 f(\tau')^2 
\Big\{ (\dd_{\tau} \dd_{\tau'} \Delta)(\tau|\tau',\tau')_n^2 +2 \omega_0(\tau')^2 (\dd_{\tau} \Delta)(\tau|\tau',\tau')_n^2
\nonum
&+& \omega_0(\tau)^2 \omega_0(\tau')^2 \Delta(\tau|\tau',\tau')_n^2 
+ 2 \om_2(\tau',\wp)^2 (\dd_{\tau} \Delta)\tau|\tau',\tau')_{n-1}^2 
\nonum  
&+& 2 \om_0(\tau)^2 \om_2(\tau',\wp^2)
\Delta(\tau|\tau',\tau')^2_{n-1} + \om_2(\tau,\wp)^2 \om_2(\tau',\wp^2) \Delta(\tau|\tau',\tau')^2_{n-2} \Big\}\,,
\nonumber
\end{eqnarray}
for $n \geq 0$, with the understanding that coefficients (\ref{sleexp1}) with a negative index vanish.

This can be used for the expansion of the two-point function. Written in the form%
\footnote{This corrects a copying error in Eq.(71) of \cite{BonusSLE}.} 
\ba
\label{sleexp3}
T^{\text{\bsle}}_p(\tau)T^{\text{\bsle}}_p(\tau')^* \is 
\frac{\sqrt{J_p(\tau) J_p(\tau')}}{2 {\cal E}_p^{\text{\bsle}}}
 \bigg(\frac{J_p(\tau,\tau_0) - i \mcE_p^{\text{\bsle}} \Delta_p(\tau,\tau_0)}{J_p(\tau,\tau_0) + i \mcE_p^{\text{\bsle}} \Delta_p(\tau,\tau_0) }\bigg)^{1/2}
 \nonum
&\times& \bigg(\frac{J_p(\tau',\tau_0) + i \mcE_p^{\text{\bsle}} \Delta_p(\tau',\tau_0)}{J_p(\tau',\tau_0) - i \mcE_p^{\text{\bsle}} \Delta_p(\tau',\tau_0) }\bigg)^{1/2},
\ea
with $J_p(\tau) = J_p(\tau,\tau)$, 
the expansions (\ref{sleexp0}), (\ref{sleexp2}) will give rise to a convergent expansion 
in powers of $|p|$. For ${\cal E}_p^{\text{\bsle}}$ the structure is slightly different depending on whether or not $\varepsilon_0 =0$. Omitting the $\wp_i$ dependence of the 
coefficients one has  
\be
\label{sleexp4}
{\cal E}_p^{\text{\bsle}} = \left\{
\begin{array}{ll} 
\displaystyle{	
\varepsilon_0 + \frac{\varepsilon_1^2}{2 \varepsilon_0} |p|^2 
- \frac{\varepsilon_1^4 - 4 \varepsilon_0^2 \varepsilon_2^2}{8 \varepsilon_0^3} |p|^4 
+ O(|p|^6) }\,, & \varepsilon_0 >0 \,,
\\[4mm]
\displaystyle{
\varepsilon_1 |p| + \frac{\varepsilon_2^2}{2 \varepsilon_1} |p|^3 
- \frac{\varepsilon_2^4 - 4 \varepsilon_1^2 \varepsilon_3^2}{8 \varepsilon_1^3} |p|^5 
+ O(|p|^7)} \,, & \varepsilon_0 = 0\,.
\end{array} \right. 
\ee 
For reasons that will become clear shortly the two cases will be referred to as
``massive'' and ``massless'', respectively. The expansion of (\ref{sleexp3}) to  
subleading order takes the following form
\medskip 

{\bf Massive:} $\varepsilon_0 >0$ 
\begin{align}
\label{sleexp5} 
&T^{\text{\bsle}}_p(\tau)T^{\text{\bsle}}_p(\tau')^* = \frac{\sqrt{J_0(\tau) J_0(\tau')}}{2 \varepsilon_0} \Big\{ 1 + \frac{|p|^2}{2} \Big( - \frac{\varepsilon_1^2}{\varepsilon_0^2}
 + \frac{J_1(\tau,\wp)}{J_0(\tau)} + \frac{J_1(\tau',\wp)}{J_0(\tau')} \Big)
\nonum
& \hspace{6cm} - i |p|^2 \big( \Upsilon_1(\tau,\wp) - \Upsilon_1(\tau',\wp) \big) + O(|p|^4) \Big\}\,,
\\[2mm]
& \Upsilon_1(\tau,\wp) = \frac{2 \varepsilon_0^2 \big( \Delta_1(\tau,\tau_0;\wp) J_0(\tau,\tau_0) - 
	J_1(\tau, \tau_0;\wp) \Delta_0(\tau,\tau_0) \big)+ \varepsilon_1^2 \Delta_0(\tau,\tau_0)
	J_0(\tau,\tau_0)}{2\varepsilon_0^3 \Delta_0(\tau,\tau_0)^2 + 2 \varepsilon_0 J_0(\tau,\tau_0)^2} \,.
\nonumber
\end{align}

{\bf Massless:} $\varepsilon_0 =0$ and $\varepsilon_1>0$.  
\begin{align}
\label{sleexp6} 
& T^{\text{\bsle}}_p(\tau)T^{\text{\bsle}}_p(\tau')^* = \frac{\sqrt{J_0(\tau) J_0(\tau')}}{2 \varepsilon_1 |p|} \Big\{ 1 - i |p| \varepsilon_1 
\Big( \frac{\Delta_0(\tau,\tau_0)}{J_0(\tau,\tau_0)} -\frac{\Delta_0(\tau,\tau_0)}{J_0(\tau,\tau_0)} \Big) + O(|p|^2)\Big\}.
\end{align}

The massive case corresponds to nonzero $\omega_0(\tau)$.
Even the lowest order commutator function $\Delta_0(\tau,\tau')$ can then in general 
no longer be found in closed form. All other aspects of the 
expansions are however explicitly computable in terms of $\Delta_0$:
the $\Delta_n$'s via (\ref{Deltaexp1}), and then the $J_n$ as well as $\varepsilon_n$ 
in terms of the $\Delta_n$ via (\ref{sleexp2}).  

\medskip

{\bf Two-point function of massless SLE.} The massless case 
corresponds to $\omega_0(\tau) \equiv 0$. The lowest order wave equation 
in (\ref{Srec1}) is then trivially soluble: $S_0(\tau) = w_0(\tau\!-\!\tau_0) + z_0$, with
$w_0 z_0^* - w_0^* z_0 = -i$. The coefficients of the commutator 
function are explicitly known 
\begin{eqnarray}
\label{massless1} 
\Delta_0(\tau',\tau) &\! =\! & \tau' - \tau\,, 
\nonumber \\[1.5mm]
\Delta_1(\tau',\tau) &\! =\! & \int_{\tau}^{\tau'} \! ds \,
\omega_2(s,\wp)^2\,(\tau -s)(\tau'-s)\,,
\end{eqnarray}
etc.. This entails $\varepsilon_0=0$, and 
\begin{equation}
\label{massless2}
\varepsilon_1^2 = \frac{1}{4} 
\int\!d\tau f(\tau)^2 \int\! d\tau' f(\tau')^2 \om_2(\tau',\wp)^2\,,
\quad 
J_0(\tau,\tau_0)= \frac{1}{2} \int\!d\tau f(\tau)^2 \,.
\end{equation}
So far we used a generic quadratic dispersion (\ref{dispersion}). For 
illustration we now specialize to the massless Bianchi I dispersion 
(\ref{Bdispersion}), i.e.~to $\om_0 \equiv 0$, $\om_2^{ij}(\tau)
= \wg(\tau) \wg^{ij}(\tau)$. Inserted into (\ref{sleexp6}) this yields
\begin{eqnarray}
\label{massless3} 
&& T_p^{\text{\bsle}}(\tau) T_p^{\text{\bsle}}(\tau')^* = \frac{1}{2\sqrt{p_i Q^{ij} p_j}} - 
\frac{i}{2} (\tau\!-\! \tau') + O(|p|)\,,
\nonum
&& Q^{ij}:=\frac{\int\! d\tau f(\tau)^2 \wg(\tau) \wg^{ij}(\tau)}{\int \!d\tau f(\tau)^2}\,. 
\end{eqnarray}
Performing the inverse spatial Fourier transform leads to the announced position 
space decay (\ref{sleIR}). For a general quadratic dispersion the same holds
with $\om_2^{ij}(\tau)$ replacing $\wg(\tau) \wg^{ij}(\tau)$ in the definition 
of $Q^{ij}$.  

\pagebreak[3] 
\noindent
{\bf Remarks.} 

(i) The massless \bsle have the {\it same} $O(1/|p|)$ type 
behavior for {\it all} Bianchi I geometries and the inverse Fourier transform 
is always infrared finite. This generalizes an analogous finding in 
\cite{BonusSLE} for the SLE on \FL spacetimes. It is a genuinely quantum 
phenomenon, which in general does not reflect the behavior of otherwise motivated
classical solutions.     

(ii) In  the massless case minimization of ${\cal E}_p[T]$ and 
expansion in $|p|^2$ do not commute. While the chosen fiducial solution 
has a regular $p_i \rightarrow 0$ limit, the associated \bsle solution 
is singular as $p_i \ra 0$. The independence of the \bsle solution from the choice of 
fiducial solution is crucial for the result.  

(iii) In the massless case the action (\ref{action1}) has the  
the shift symmetry, $\phi(\tau,x) \mapsto \phi(\tau, x) + {\rm const}$.
This symmetry turns out to be spontaneously broken for $d \geq 2$, 
in parallel to the massless free field in Minkowski space. A proof 
could be based on Swieca's Noether charge criterion \cite{Swieca,Lopusbook}.

In arriving at (\ref{massless3}) we used the expressions (\ref{ssle10}) for the \bsle 
two-point function in terms of the commutator function. It is instructive to 
rederive the result starting from the expression (\ref{sle5}) of the \bsle solution 
in terms of a fiducial solution. With $c_1$, $c_2$ from (\ref{sle4}) the 
basic expansions are 
\be
\label{fidexp1}
c_1 = \sum_{n \geq 0} c_{1,n} |p|^{2n} \,, \quad 
c_2 = \sum_{n \geq 0}c_{2,n} |p|^{2n} \,,\quad 1- \frac{|c_2(p)|^2}{c_1(p)^2} = \al_0+\sum_{n=1}^\infty \al_n |p|^{2n}\,,
\ee 
with radius of convergence $|p| < |p|_*$. For simplicity, we suppress the residual 
$\wp_i$ dependence of the coefficients in the notation, 
$c_{1,n} = c_{1,n}(\wp)$,  $c_{2,n} = c_{2,n}(\wp)$, $n \geq 1$.  
The $\alpha_n$ are of course combinations of the $c_{1,n}, c_{2,n}$; 
due to the form of (\ref{sle3}) only they will enter the $\mu_p,\lb_p$
expansion. A nonzero $\alpha_0$ turns out to correspond to the massive case
while $\alpha_0 = 0$ for $\om_0 \equiv 0$. This entails a  
qualitatively different behavior in both cases.

We focus again on the massless case. Then $S_0(\tau) 
= w_0(\tau\!-\!\tau_0) + z_0$, $w_0 z_0^* - w_0^* z_0 = -i$, as noted before.
For the basic expansion coefficients one has 
\ba 
\label{massless4} 
c_{1,0} \is \frac{|w_0|^2}{2} \int\! d\tau f(\tau)^2 \,,
\nonum
c_{1,1} \is \frac{1}{2} \int\! f(\tau)^2 \big\{ 2 
\Re( w_0^* \dd_{\tau}S_1) + \om_2(\tau,\wp)^2 | w_0(\tau\! -\! \tau_0) + z_0|^2 
\big\}\,,
\nonum
c_{2,0} \is \frac{w_0^2}{2} \int\! d\tau f(\tau)^2 \,,
\nonum
c_{2,1} \is \frac{1}{2} \int\! f(\tau)^2 \big\{ 2 
w_0 \dd_{\tau}S_1 + \om_2(\tau,\wp)^2(w_0(\tau\! -\! \tau_0) + z_0)^2 
\big\}\,,
\ea
where the explicit form of $S_1$ will not be needed. This results in 
\be
\label{massless5} 
\alpha_0 =0\,, \quad \alpha_1 = \frac{\displaystyle{\int\! d\tau f(\tau)^2 f(\tau)^2 \om_2(\tau,\wp)^2}}{\displaystyle{|w_0|^4 \int\! d\tau f(\tau)^2}}\,. 
\ee
Using (\ref{massless5}) one finds to subleading order 
\begin{eqnarray}
\label{massless6}
\mu_p\is \frac{1}{\alpha_1^{1/4} \sqrt{2 |p|}}\Big( 1 - \frac{1}{2} \alpha_1^{1/2} |p| +
O(|p|^2) \Big) \,, 
\nonum 
\lb_p \is - \frac{w_0^*}{w_0}\frac{1}{\alpha_1^{1/4} \sqrt{2 |p|}}\Big( 1 + \frac{1}{2} \alpha_1^{1/2} |p| +
O(|p|^2) \Big) \,.
\end{eqnarray}
Finally, (\ref{sle5}) evaluates to  
\begin{eqnarray}
\label{massless7}
T_p^{\text{\bsle}}(\tau)\is \lb_p \,S_p(\tau)+\mu_p \,S_p(\tau)^\ast
\\[2mm]
\is \frac{-1}{\sqrt{2 |p|}\alpha_1^{1/4}w_0}\Big\{ i + \frac{1}{2} 
|p| \alpha_1^{1/2} \big[ 2 |w_0|^2 (\tau - \tau_0) + w_0^* z_0 + w_0 z_0^* \big] 
+ O(|p|^2) \Big\}\,.
\nonumber
\end{eqnarray}
Interestingly, the leading IR behavior of the massless \bsle solution 
is  constant ($\tau$ independent) for 
any Bianchi I geometry. In isotropic \FL spacetimes this reflects     
the expected freeze-out of the oscillatory behavior on scales much larger 
than the Hubble radius. The result (\ref{massless7}) shows that anisotropy 
does not interfere with the freeze-out.  The universality of the $|p|^{-1/2}$ 
behavior is surprising, as is the bounded anisotropy modulation 
contained in the coefficient $(4\alpha_1)^{-1/4}$. The result 
(\ref{massless7}) could not have been obtained by means of the traditional 
adiabatic iteration, which is incurably singular at small momenta. Finally,
note that (\ref{massless7}) depends on the parameters $w_0,z_0$ of the fiducial 
solution. In the two-point function, however, these drop out and one recovers (\ref{massless3}). 


\section{Ultraviolet behavior of \bsle states} 
\label{sec5}

The ultraviolet behavior of Hadamard two-point functions is generally captured 
by the Hadamard expansion of the coefficient functions of its singular parts,
see e.g.~\cite{Wavebook,heatkoffdiag1}. For spatially homogeneous spacetimes
this can be replaced by a (considerably simpler) expansion in inverse powers of the 
modulus of the spatial momentum, see \cite{Pirk,bianchiAd1,BianchiHad1}. 
Here we show that an asymptotic expansion of this type with specific,
recursively determined coefficients is in fact {\it equivalent} to the Hadamard property.
Moreover, a closed, non-recursive formula for these coefficients is presented.

\subsection{Solution of the Gelfand-Dickey recursion} 
\label{subsec5a}

We return to the basic wave equation (\ref{wave2}) but aim now at an expansion valid 
for large $|p|$. Throughout this subsection we write $T_p(\tau)$ for a (Wronskian 
normalized) solution of (\ref{wave2}). As such it is uniquely determined by its 
modulus and we can write 
\be
\label{GD0} 
T_p(\tau) = |T_p(\tau)| \exp\Big\{\!\!-\frac{i}{2} \int_{\tau_0}^{\tau} \! ds\, \frac{1}{|T_p(s)|^2} 
\Big\}\,.
\ee
Further, the modulus-square solves the following Gelfand-Dickey equation
\be  
\label{GD1} 
2 |T_p(\tau)|^2 \,\dd_{\tau}^2 |T_p(\tau)|^2 - \big( \dd_{\tau} |T_p(\tau)|^2 \big)^2 
+ 4 \om_p(\tau)^2 |T_p(\tau)|^4 =1\,.
\ee
It is convenient to use the squared coefficient functions  in the dispersion relation, $\om_p(\tau)^2 = 
\om_0(\tau)^2 + |p|^2 \om_2(\tau,\wp) =: v(\tau) + |p|^2 w(\tau,\wp)$, where we normally suppress the 
arguments of $v,w$. Next, we search for a solution of (\ref{GD1}) via an initially formal 
series ansatz of the form  
\be 
\label{GD2}
|T_p(\tau)|^2= \frac{1}{2 \sqrt{w} |p|} \sum_{n \geq 0} (-)^n G_n(v,w) |p|^{-2n} \,.
\ee
Inserting (\ref{GD2}) into (\ref{GD1}) and comparing  powers of $1/|p|$ one finds
$G_0 =1$ and for $n \geq 1$ the nonlinear recursion \cite{BonusSLE}  
\ba 
\label{GD3} 
G_n \is \!\!\sum_{k,l\geq 0, k+l =n-1} \!\Big\{ 
\frac{1}{4} \frac{G_k}{\sqrt{w}} 
\partial_{\tau}^2 \Big( \frac{G_l}{\sqrt{w}} \Big) - \frac{1}{8} 
\partial_{\tau} \Big( \frac{G_k}{\sqrt{w}} \Big) 
\partial_{\tau} \Big( \frac{G_l}{\sqrt{w}} \Big) 
\nonum
&+& \frac{1}{2} \frac{v}{w} G_k G_l \Big\} 
-\frac{1}{2} \sum_{k,l\geq 1, k+l =n} G_k G_l\,. 
\ea
This expresses $G_n$ in terms of $G_{n-1}, \ldots, G_1$, and involves only $\dd_{\tau}$ 
differentiations. For short we shall refer to the $G_n$'s as the Gelfand-Dickey 
(GD) coefficients. In explicit expressions is convenient to write $\pp_{\tau} f=: \dot{f}$ 
for the reparameterization invariant derivative. Iteratively, a differential polynomial in 
$G_1$ arises, where
\be
\label{GD4a}
G_1 = \frac{v}{2 w} + \frac{5}{32} \frac{\dot{w}^2}{ w^3}
- \frac{1}{8} \frac{\ddot{w}}{ w^2}\,.
\ee
The terms without and with the maximal number of derivatives
are easily inferred
\be
\label{GD4b}
G_n = \frac{(2n\!-\!1)!!}{n!} G_1^n + \ldots + \frac{1}{(2\sqrt{w})^{2n-2}}
\dd_{\tau}^{2 n-2} G_1\,, \quad n\geq 2\,.
\ee
A more detailed inspection shows that the iterates are of the form 
\ba 
\label{GD5} 
G_n(v,w) \is \frac{1}{w^n}
\sum_{j=0}^{2n} \frac{g_{n,j}(v,\dot{w})}{w^j}\,, 
\nonum
g_{n,0}(v) \is \frac{(2n\!-\!1)!!}{2^n} A_{n}\big(\!-\!\dd_{\tau}^2 - v\big) \,. 
\ea
Here, $A_n$ is the conventionally normalized heat kernel coefficient
associated with the one-dimensional Schr\"{o}dinger operator 
$\!-\!\dd_{\tau}^2 - v$. Our normalizations are such that 
$A_n(-\dd_{\tau}^2 +v) = (-)^n v^n/n! \,+ $ derivative terms. 
For $j \geq 1$, the key feature is that $g_{n,j}$
is a polynomial in the derivatives of $w$ only, specifically one of degree $j$. 
The recursion (\ref{GD3}) is easily programmed and 
allows one to read off the $g_{n,j}$ to moderately high orders. 
To low orders: $g_{0,0} =1$ and
\ba 
\label{GD7}
g_{1,0} \is \frac{1}{2} v\,,  \quad 
g_{1,1} = \frac{1}{8} \ddot{w}\,,\quad 
g_{1,2} = \frac{5}{32} \dot{w}^2\,.
\nonum
g_{2,0} \is \frac{1}{8} \big( v^2 + 3 \ddot{v} \big) \,, \quad 
g_{2,1} = -\frac{5}{16} (v \ddot{w} + \dot{v} \dot{w} ) - \frac{1}{32} w^{(4)}\,, 
\nonum
g_{2,2} \is \frac{35}{64} v \dot{w}^2 + \frac{7}{32} \dot w \dddot{w} + \frac{21}{128} 
\ddot{w}^2\,,
\nonum
g_{2,3} \is - \frac{231}{256} \dot{w}^2 \ddot{w} \,, \quad 
g_{2,4} = \frac{1155}{2048} \dot{w}^4\,.
\ea 
One can also convert the recursion (\ref{GD3}) into one directly for the 
tuples $(g_{n,1}, \ldots, g_{n,2n})$, thereby streamlining the task of computing 
them. Improving on the recursion one has: 

\begin{proposition} \ \label{GDcoeff} Let $G_n$, $n \geq 1$, be the recursively 
	defined coefficients 
(\ref{GD3}) with $G_1$ given by (\ref{GD4a}). Then: 
\begin{itemize} 
\item[(a)]
Up to a normalization the $G_n$'s are the one-dimensional heat kernel
coefficients for the Schr\"{o}dinger operator $-(w^{-1/2}\dd_{\tau})^2 - 2 G_1$, i.e.
\ba
\label{nonrec2} 
G_n(v,w) \is G_n(2 G_1, 1) \Big|_{\dd_{\tau} \mapsto w^{-1/2} \dd_{\tau}} = 
g_{n,0}(2 G_1) \Big|_{\dd_{\tau} \mapsto w^{-1/2} \dd_{\tau}} 
\nonum
\is \frac{(2n\!-\!1)!!}{2^n} A_{n}\big(\!-\!(w^{-1/2} \dd_{\tau})^2 - 2G_1\big) \,, 
\quad n \geq 2\,.
\ea
By expansion (\ref{GD5}) it follows that for all $n \geq 2$ the
$g_{n,j}(v,\dot{w}), j=1,\ldots,n$, are uniquely determined by $g_{n,0}(v)$. 
\item[(b)] Explicitly: 
\ba 
\label{nonrec3} 
\nspace G_n \is \frac{(2n\!-\!1)!!}{2^n} \sum_{j=0}^n { n\!+\!1/2 \choose n\!-\!j} 
\frac{(-)^j (2 j)!}{4^j j! (j\!+\!n)!}\sum_{p=1}^{n} 2^{p}
\nonum
\nspace &\times&  \nspace
\sum_{ \begin{array}{c} \scriptstyle k_1, \ldots, k_p \geq 0 \\[-1.5mm]
    \scriptstyle k_1 + ...+ k_p = 2(n-p) \end{array} }
\nspace {}_{n+j} C_{k_1 \ldots k_p} 
(w^{-1/2} \dd_{\tau})^{k_1} G_1 \ldots (w^{-1/2}\dd_{\tau})^{k_p} G_1\,, 
\ea 
where the coefficients are for $p=1,\ldots, n$ given by
\ba
\label{nonrec4}
 \nspace     {}_{n} C_{k_1 \ldots k_p} = \nspace \sum_{ \begin{array}{c}
     \scriptstyle 0 \leq m_1 \leq m_2 ... \leq m_p \leq n -p \\[-1.5mm]
     \scriptstyle 2 m_i \geq k_1 + ... + k_i, i=1, ..., p \end{array}}
     \nspace { 2 m_1 \choose k_1} { 2 m_2 - k_1 \choose k_2} \cdots
      { 2 m_p - k_1 -... - k_{p-1} \choose k_p} .
\ea 
\end{itemize} 
\end{proposition}

{\it Proof:} 
(a) We define $T_p =: w^{-1/4} \tilde{T}_p$ and replace iterated $\dd_{\tau}$ derivatives 
by iterated $w^{-1/2} \dd_{\tau}$ derivatives.  Since $w = w(\tau,\wp)$ is momentum
dependent this is slightly unusual, but for fixed $\wp_i$ the transformation is well-defined. A straightforward computation shows that $\tilde{T}_p$ satisfies  
\ba
\label{nonrec1} 
&& \big\{ (w^{-1/2} \dd_{\tau})^2  + 2 G_1 + |p|^2\big\} \tilde{T}_p =0\,, 
\nonum
&& (w^{-1/2} \dd_{\tau} \tilde{T}_p) \tilde{T}_p^* - (w^{-1/2} \dd_{\tau} \tilde{T}_p)^* \tilde{T}_p =-i\,.
\ea
Note that $|p|^2$ now occurs with constant coefficient like a resolvent parameter. Moreover, the wave 
operator is a one-dimensional Schr\"{o}dinger operator with potential $-2 G_1$ and derivative 
$w^{-1/2} \dd_{\tau}$. One can obtain the associated Gelfand-Dickey equation as 
before and derive an asymptotic expansion for large $|p|$. Comparing the 
result with (\ref{GD3}) gives the first and the second identity in (\ref{nonrec2}).
For one-dimensional Schr\"{o}dinger operators the coefficients arising 
in the diagonal resolvent expansion are proportional to the diagonal heat kernel coefficients, see e.g.~\cite{Avramidibook}. The signature change (Lorentzian for the 
resolvent proper and Euclidean for the heat kernel) can be attributed to a sign 
flip in the potential and by matching the normalizations one obtains the last 
equality in (\ref{nonrec2}).

(b) Several authors have derived non-recursive 
formulas for the one-dimensional diagonal heat kernel coefficients 
\cite{AvSch, Polter}. The expression in \cite{Polter}, Theorem 5.2.2, derives 
from the specialization of Polterovich's remarkable non-recursive expression 
for the general case. It requires a `normal ordering' formula for the powers 
of the differential operator in question, taken in the one-dimensional
case from \cite{Rida}. Note that the expressions for ${}_nC_{k_1,\ldots,k_p}$ in 
\cite{Rida} and \cite{Polter} contain subtle typos, which are corrected in 
(\ref{nonrec4}). Combining these results with (\ref{nonrec2}) we
obtain (\ref{nonrec3}), (\ref{nonrec4}). \qed

{\bf Remarks.}

(i) For $w\mapsto 1$, $-2G_1 \ra V$, the formula (\ref{nonrec3}) provides
a closed expression for the one-dimensional heat kernel coefficients. 
It can be programmed in {\tt Mathematica} and avoids the RAM overflow 
typically occurring in a symbolic recursive evaluation at 
fairly low orders. For example, $A_{10}$ evaluates via (\ref{nonrec3}) 
in $100$ sec on a laptop.  

(ii) Concerning typos, we note as a benchmark the coefficients (\ref{nonrec4}) for 
$n=3$: ${}_3 C_0 =3$, ${}_3 C_1 =6$, ${}_3 C_2 = 7$, ${}_3 C_3 =4$, ${}_3 C_4 =1$, 
${}_3 C_{00} =1$, ${}_3 C_{01} =4$,
${}_3 C_{02} =2$, ${}_3 C_{10} =2$, ${}_3 C_{11} =2$, ${}_3 C_{20}=1$.

(iii) For the evaluation of the GD coefficients (\ref{nonrec3}) one might 
want to separate the undifferentiated and the differentiated occurrences of $w$.
The separation amounts to evaluating the coefficients in
\begin{equation}
\label{nonrec5}
(w^{-1/2} \dd_{\tau})^n G_1 = \sum_{j =0}^{n+2}
\frac{d_{n,j}(v, \dot{w})}{w^{n/2 + 1 + j}}\,,\quad n \geq 1\,.
\end{equation}
In practice, one can find the $d_{n,j}$ coefficients by programming
the $w^{-1/2} \dd_{\tau}$ derivative in {\tt Mathematica}
and storing the results in a look-up table. For $1 \leq n \leq 20$ generation
of the look-up table takes a few minutes. If on matters of principle
an explicit formula is sought one can proceed as follows. Denoting
momentarily the left hand side of (\ref{nonrec5}) by $D_n$ one has
$D_n = w^{-1/2}\dd_{\tau} D_{n-1}$, which results in a linear `triangular' 
recursion for the $d_{n,j}$'s, which should allow one to   
obtain explicit formulas for the $d_{n,j}$.

Conceptually, the expansion (\ref{GD2}) ought to be related to the 
general Hadamard expansion on globally hyperbolic spacetimes. 
However, since the Synge function on 
Bianchi I geometries cannot be found in closed form and a distributional 
inverse Fourier transform enters, the precise relation is technically 
rather involved. Conversely, one can view  (\ref{GD2}) and Proposition 
\ref{GDcoeff} as a technically advantageous simplification of the general 
Hadamard expansion when applied to Bianchi I geometries.
Since the general Hadamard coefficients can, after formal 
Wick rotation,  be expressed in terms of the heat kernel coefficients, 
the analysis can be simplified by using a suitably complexified 
heat kernel and its asymptotic expansion. The Gelfand-Dickey 
coefficents $G_n$ can then eventually be related to the spatially off-diagonal  
Bianchi I heat kernel coefficients on a fixed time slice. We defer the 
details to a separate publication.


\subsection{All order approximants and the Hadamard property}  
\label{subsec5b}

So far the expansion (\ref{GD2}) has been treated as formal. The goal in the 
following is to show that under suitable subsidiary conditions the expansion 
can be promoted to an asymptotic expansion. Moreover, the validity 
of this asymptotic expansion links back precisely to the Hadamard property.
We begin with the following 

\begin{definition}
A smooth function $A_p^{(N)}: J \ra \bbC$, where $J\subseteq \bbR$ is an open interval, is called an 
{\it approximant of order $N \in \bbN$} if 
\ba
\label{approx0}
&& |A_p^{(N)}(\tau)|^2 = \frac{1}{2 \sqrt{w} |p|} 
\Big\{ 1 + \sum_{n=1}^{N} (-)^n G_n(\tau,\wp)|p|^{-2n} + 
O(|p|^{- 2N -2}) \Big\}\,,
\ea
uniformly in every compact $[\tau_i, \tau_f]\subseteq J$
and $N$ is the largest integer for which this holds  (with the understanding ``$N=\infty$'' if \eqref{approx0} holds for every $N\in \bbN$). Here, $G_1, \ldots, G_N$, 
are the GD coefficients from Proposition \ref{GDcoeff}.
\end{definition}

By remark (iv) to Theorem \ref{thadmain} the adiabatic iterates of order $N$ 
are approximants of order $N$ in this sense. Computationally much 
spurious data are carried along by the iteration, however. The WKB approximants in 
Section IV.A of \cite{BonusSLE} provide another, more direct construction 
principle. The characterization from  Proposition \ref{GDcoeff} is arguably the 
most concise way of constructing approximants, with no `junk' carried along.

Given such an $A_p^{(N)}(\tau)$ we define 
\be 
\label{approx1} 
W_p^{(N)}(\tau) := |A_p^{(N)}(\tau)| \exp \Big\{ \!\!- \frac{i}{2}
\int_{\tau_0}^{\tau} \! ds \frac{1}{|A_p^{(N)}(s)|^2} \Big\}\,.
\ee
It satisfies
\ba
\label{approx2}
&& \dd_{\tau} W_p^{(N)} [W_p^{(N)}(\tau)]^* - 
[\dd_{\tau} W_p^{(N)}]^* W_p^{(N)}(\tau) = - i\,,
\nonum
&& \big[ \dd_{\tau}^2 + \om_p^2 \big] W_p^{(N)} = 
\bigg[ \om_p^2 + \frac{\dd_{\tau}^2 |A_p^{(N)}|}{|A_p^{(N)}|} 
- \frac{1}{4 |A_p^{(N)}|^4} 
\bigg] W_p^{(N)} \,.
\ea 
Clearly, $W_p^{(N)}$ is not a solution of the homogeneous wave equation. 
However, for large $|p|$ the right hand side is small
\be
\label{approx3}
\om_p^2 + \frac{\dd_{\tau}^2 |A_p^{(N)}|}{|A_p^{(N)}|} 
- \frac{1}{4 |A_p^{(N)}|^4} = O(|p|^{-2N} )\,.
\ee
To see this, we note that the Gelfand-Dickey equation (\ref{GD1}) can be rewritten as
\be 
\label{approx4} 
\om_p^2 + \frac{ \dd_{\tau}^2 |T_p|}{ |T_p|}- \frac{1}{4 |T_p|^4 }  =0 \,,
\ee
where an alternative form of the derivative term is $\dd_{\tau} |T_p|^2/(2 |T_p|^2) 
- (\dd_{\tau} |T_p|^2)^2/(4 |T_p|^4)$.  Applying the latter form to 
$|A_p^{(N)}|^2$ the left hand side of (\ref{approx3}) evidently has an 
expansion   in powers of $1/|p|^2$. The coefficient of $|p|^0$ is
just the defining relation (\ref{GD4a}) of $G_1$ and thus vanishes. For 
$n=1, \ldots, N$, the coefficient of $|p|^{-2n+2}$ is a   
a differential polynomial in $G_1, \ldots, G_n$ with coefficients
built from $1/\sqrt{w}$ and its derivatives.  By construction, this
relation is equivalent to (\ref{GD3}), (\ref{GD4b}); so all the 
coefficients of $|p|^{-2n}, n =0, \ldots, N\!-\!1$ vanish. 
This yields (\ref{approx3}).

An estimate of the form (\ref{approx3}) is the core of the proof of Theorem \ref{thad1}.
The other aspects of the proof carry over straightforwardly to 
an approximant of order $N$. We note this as 

\begin{proposition} \label{SNviaapprox}
Let $A_p^{(N)}$ be an approximant of order $N\geq 1 $ and $W_p^{(N)}$ the 	
associated WKB approximant (\ref{approx1}). Then, there exists an exact 
Wronskian normalized solution $S_p^{(N)}$ of the wave equation such that 
\ba
\label{approx5}
S_p^{(N)}(\tau) \is W_p^{(N)}(\tau) \big[ 1 + O(|p|^{-2N})\big]\,,
\nonum
\dd_{\tau} S_p^{(N)}(\tau) \is \dd_{\tau} W_p^{(N)}(\tau) \big[ 1 + O(|p|^{-2N})\big]\,,
\ea
uniformly for $\tau \in [\tau_i, \tau_f]$ as $|p| \ra \infty$. 
\end{proposition} 	
	
Similarly as in Corollary \ref{corad1} this implies 
\be 
\label{approx6} 
\dd_{\tau}^{\ell} S_p^{(N)}(\tau) = O(|p|^{\ell -1/2})\,,
\quad \ell \in \bbN_0\,, N\in \bbN\,, 
\ee 	
uniformly for $\tau \in [\tau_i,\tau_f]$, where for $\ell \geq 2$ the differential 
equation enters.  	
	
For a generic quadratic dispersion \eqref{dispersion} with $\omega_0^2=v$ and $\omega_2^2=w$,	the main result of this section is		
\begin{theorem} \label{HadvsGD} 
Let $T_p$ be a Wronskian normalized solution of $[\dd_{\tau}^2 + v + |p|^2 w] T_p =0$
and let $\varpi$ be the Wightman function (\ref{Bcorr1}) built from it. 
Then $\varpi$ is associated with a Hadamard state if and only if $|T_p(\tau)|^2$ 
admits an asymptotic expansion of the form 
\be 
\label{HadvsGD1} 
|T_p(\tau)|^2 \asymp \frac{1}{2 \sqrt{w} |p|} \Big\{ 
1 + \sum_{n \geq 1} (-)^n G_n(\tau,\wp) |p|^{-2n} \Big\}\,,
\ee
where $G_n, n \in \bbN$, are the GD coefficients. 
\end{theorem} 	
	
{\bf Remark.} This is a generalization of the corresponding result for \FL geometries
stated in \cite{FLFRG}, which we prove here in full.   	
	
{\itshape Proof of Theorem \ref{HadvsGD}:} For the `if' part, let $T_p$ be a solution whose modulus square 
has the asymptotic expansion (\ref{HadvsGD1}). Since any pair of Wronskian 
normalized solutions of the wave equation are related by a Bogoliubov transformation
we can write 
\be 
\label{HadvsGD2} 
T_p(\tau) = \alpha_p S_p^{(N)}(\tau) + \beta_p S_p^{(N)}(\tau)^*\,, 
\quad  |\alpha_p|^2 - |\beta_p|^2 =1\,,
\ee
for any of the exact solutions from Proposition \ref{SNviaapprox}. 
Then 
\ba
\label{HadvsGD3}
|T_p(\tau)|^2 \is |A_p^{(N)}(\tau)|^2 \,B_p^{(N)}(\tau) (1 + O(|p|^{-2N}))\,,
\\[2mm]
B_p^{(N)}(\tau) &:=&  |\alpha_p|^2 + |\beta_p|^2 + 2 
|\alpha_p| |\beta_p| \cos\Big[ {\rm Arg} \alpha_p\! - \!{\rm Arg} \beta_p 
-\! \int_{\tau_0}^{\tau} \! ds \frac{1}{|A_p^{(N)}(s)|^2} \Big].
\nonumber
\ea 
Inserting the defining relation (\ref{approx0}) gives 
\ba
\label{HadvsGD4}
&& \Big( 1 + \sum_{n=1}^N (-)^n G_n |p|^{-2n} \Big) 
\Big(1 - B_p^{(N)} \big(1\!+\! O(|p|^{-2N}) \big) \Big) 
\nonum
&& \sspace = B_p^{(N)} \big(1\!+\! O(|p|^{-2N}) \big) O(|p|^{-2N-2})\,.
\ea
Solving for the combination $1 - B_p^{(N)} (1\!+\! O(|p|^{-2N}))$
yields $1 - B_p^{(N)} (1\!+\! O(|p|^{-2N})) = O(|p|^{-2N -2})$,
and hence $B_p^{(N)} = 1 + O(|p|^{-2N})$. On the other hand, 
the definition of $B_p^{(N)}$ implies 
\be 
\label{HadvsGD5}
1 - |\beta_p| \leq \sqrt{ 1 + |\beta_p|^2} - |\beta_p|
\leq B_p^{(N)}(\tau) \leq \sqrt{ 1 + |\beta_p|^2} + |\beta_p| 
\leq 1 + 2 |\beta_p|\,.
\ee
Hence 
\be 
\label{HadvsGD6} 
|\alpha_p| = 1 + O(|p|^{-2N}), \quad  |\beta_p| = O(|p|^{-2N})\,. 
\ee	
Inserting into (\ref{HadvsGD2})	and using (\ref{approx5}), as well as 
$W_p^{(N)} = O(|p|^{-1/2})$ gives $T_p(\tau) = S_p^{(N)}(\tau) + O(|p|^{-2N -1/2})$.
Similarly, using (\ref{HadvsGD6}) in the derivatives of (\ref{HadvsGD2})
in combination with (\ref{approx6}) results in 
\be 
\label{HadvsGD7}
\dd_{\tau}^{\ell}T_p(\tau) = \dd_{\tau}^{\ell} S_p^{(N)}(\tau) + O(|p|^{\ell-2N -1/2})\,,
\quad \ell \in \bbN_0\,,\;N \in \bbN\,.
\ee
This is the counterpart of the crucial estimate (\ref{bsad1}) in Lemma \ref{lmbsad}. 
Re-examination of the proof in Section \ref{subsec3b} shows that subject to this estimate
the solution $T_p(\tau)$ is associated with a Hadamard state. This shows
the `if' direction of the theorem.

For the converse, we first show that the modulus square of the \bsle mode 
$T_p^{\text \bsle}(\tau)$ function  has an asymptotic expansion of the asserted 
form. By Eq.~(\ref{bsad1}) and Corollary \ref{corad1} we have 
\be 
\label{HadvsGD8}
|T_p^{\text \bsle}(\tau)|^2 = |S_p^{(N)}(\tau)|^2 \big(1 + O(|p|^{-2N})\big)\,, 
\ee
for all $N \in \bbN$.  Since adiabatic vacua of order $N$ are by Appendix \ref{AppA} 
also approximants of order $N$, the  $|S_p^{(N)}(\tau)|^2$ can be 
replaced with $|A_p^{(N)}(\tau)|^2(1+ O(|p|^{-2N -2}))$. Thus 
$|T_p^{\text \bsle}(\tau)|^2$ has an expansion of the form 
(\ref{approx0}) for any $N \in \bbN$, with a remainder that 
is $O(|p|^{-2N -2})$. Hence $|T_p^{\text \bsle}(\tau)|^2$ has an 
asymptotic expansion of the form (\ref{HadvsGD1}). Next, consider 
an arbitrary Hadamard mode function $T_p^{\rm H}(\tau)$. It can be expressed as
\be 
\label{HadvsGD9}
T^{\rm H}_p(\tau) = \alpha_p T_p^{\text \bsle}(\tau) + \beta_p T_p^{\text \bsle}(\tau)^*\,, 
\quad  |\alpha_p|^2 - |\beta_p|^2 =1\,.
\ee
From the proof of Theorem \ref{thHad}, we have $\text{WF}(\varpi^\text{\bsle}
\!-\!\varpi^{\rm H})=\emptyset$, where $\varpi^\text{\bsle},
 \varpi^{\rm H}$ are the associated  two-point functions (\ref{Bcorr1}). It follows that the difference between the two-point functions must be smooth on the product manifold.
In particular, arbitrary covariant derivatives in either spacetime coordinate 
exist and are finite on the diagonal. (This feature is also visible from the 
truncated Hadamard expansion of the regular part). Specializing 
to $\tau = \tau'$ we can consider (initially distributional) 
spatial derivatives $\dd_x^{\alpha} = \dd^{a_1}_{i_{1}}\ldots \dd^{a_l}_{i_l}$ 
of the difference two-point function. Evaluated on the spatial diagonal $x = x'$ this produces 
a momentum integral with integrand $(|T^{\rm H}_p(\tau)|^2- |T_p^{\text \bsle}(\tau)|^2 ) 
p_{i_1}^{a_1} \ldots p_{i_l}^{a_l}$. In order for these integrals to be 
finite for all $a_1+ \ldots + a_l \in \bbN$, the difference 
$|T^{\rm H}_p(\tau)|^2- |T_p^{\text \bsle}(\tau)|^2$ must decay 
faster than any power. Hence also $|T_p^{\rm H}(\tau)|^2$ has an asymptotic 
expansion of the form  (\ref{HadvsGD1}).
\qed

\section{Conclusions} 

Bianchi I spacetimes are of considerable interest in primordial cosmology.
The \bsle states constructed here provide a matching class of primordial 
vacua that allow one to study quantum field theories in an anisotropic 
pre-inflationary phase. As shown, the \bsle are exact Hadamard states and have 
systematically computable infrared as well as ultraviolet expansions.
In particular, for massless \bsle the universal long distance asymptotics 
found is relevant for cosmological perturbations and their power spectrum.

A natural extension is to other homogeneous spacetimes in the Bianchi classification.
For the solvable Bianchi II-VII groups the relevant harmonic analysis (generalizing the 
spatial Fourier transform) was developed in \cite{Bianchiharmonic}. For type 
VIII, IX the harmonic analysis is known. In a quantum field theoretical 
context one would want the wave equation to arise from a symmetry reduced action 
principle. This is known to be case for the I,II,${\rm VI}_0$,${\rm VII}_0$,VIII,IX 
types (collectively known as Bianchi A, with an unimodular group of isometries). 
A generalization to such backgrounds would seem relevant and feasible.  

The primordial power spectrum in the \bsle vacua can for a
realistic expansion history be computed numerically along
the lines of \cite{BonusSLE} (see also \cite{LQCSLE1} for a preferred
\FL power law). Existing numerical codes \cite{isotropy} can then be used to
computationally propagate the primordial spectrum to the
surface of last scattering. A near exponential intermediate
expansion will tend to wash out primordial anisotropies;
the quantitative aspects of this should be worth exploring
to ensure viability of the \bsle as primordial vacua.
 
Our expansion in inverse powers of the spatial momentum scale shortcuts 
the general Hadamard expansion on globally hyperbolic spacetimes. 
After a formal Wick rotation the Hadamard coefficients 
can be related to the heat kernel coefficients, see e.g.~\cite{heatkoffdiag1}.
Investigating this interplay for the Bianchi I geometries leads
to a concise expression for the spatially off diagonal heat kernel 
coefficients, on which we will report elsewhere. Beyond the formal level one 
can search for a suitably complexified diffusion kernel coding much the 
same information as the resolvent kernel, i.e.~as an associated Green's function. 
The enhanced structural and computational control over the Bianchi I Green's 
functions renders them a good test bed for these developments. 

Finally, we mention that pseudo-Riemannian geometries and the associated 
wave operators admit a `contraction' (the Anti-Newtonian limit, see 
\cite{SvsCtensor}) which resembles their Bianchi I versions. Several of the 
results reported here will `lift' to this no longer symmetry restricted 
setting.   

\newpage

\appendix

\renewcommand{\theequation}{\thesection.\arabic{equation}}

\section{Adiabatic vacuum states on Bianchi I}
\label{AppA}

The concept of adiabatic vacua was first introduced for free scalar fields on \FL spacetimes in \cite{Parker1} with the aim of providing a definition of a `physical particle' in an expanding spacetime with `slowly varying' scale factor. The definition of adiabatic vacua was subsequently refined and placed on firm mathematical ground in \cite{Luders}. While adiabatic states do in general {\it not} have the Hadamard property, they can serve as a conduit to establish the existence of states locally indistinguishable from Hadamard states \cite{JunkerS}. Adiabatic vacua on Bianchi I spacetimes were considered in \cite{bianchiAd1, bianchiAd2}, but a detailed analysis of convergence and asymptotics of the adiabatic iteration procedure, and the ensued construction of adiabatic vacuum states does not appear to have been previously considered. We therefore present such an analysis here, streamlined towards a
large momentum asymptotic expansion, which we also link to the Gelfand-Dickey 
expansion from Section \ref{sec5}. Throughout this appendix we take $\om_p(\tau)$ to  be 
a general quadratic dispersion relation in the sense of Definition \ref{quaddispersion}.


\subsection{Adiabatic frequency iteration}
\label{AppA1}

We analyze here the adiabatic iteration procedure, which will be used in Section \ref{secAdvac} to define adiabatic vacuum states. Any solution of 
\be
\label{ad0} 
\big[ \partial_{\tau}^2 + \omega_p(\tau)^2 ] S_p(\tau) =0\,,\quad 
(\partial_{\tau} S_p \,S_p^*)(\tau) - (\partial_{\tau} S_p^* \,S_p)(\tau) = -i \,,
\ee
with $\omega_p(\tau)=[\omega_0(\tau)^2 + |p|^2 \omega_2(\tau,\wp)^2]^{1/2} $  a quadratic dispersion in the sense of Definition \ref{quaddispersion}, may be written in a WKB form as 
\begin{eqnarray}\label{ad1}
S_p(\tau)\is \frac{1}{\sqrt{2\Omega_p(\tau)}}\exp\bigg\{-i \int^\tau_{\tau_0}\!\!ds\,\Omega_p(s)\bigg\}\,,
\end{eqnarray}
up to a choice of $\tau_0$. This implies that  $\Omega_p$ satisfies 
\begin{eqnarray}\label{ad2}
\Omega_p^2\is \omega_p^2+\frac{1}{16}\bigg(\frac{\pp_\tau\Omega_p^2}{\Omega_p^2}\bigg)^2-\frac{1}{4}\frac{\pp_\tau^2 \Omega_p^2}{\Omega_p^2}\,,
\end{eqnarray}
where we have suppressed the $\tau$-dependence for brevity. 
Since \eqref{ad2} cannot in general be solved exactly, one pursues an iterative solution scheme for $N\in \bbN_0$
\begin{eqnarray}\label{iter1}
\Omega_p^{(N+1)}(\tau)^2\is \omega_p(\tau)^2+\frac{1}{16}\bigg(\frac{\pp_\tau\Omega_p^{(N)}(\tau)^2}{\Omega_p^{(N)}(\tau)^2}\bigg)^2-\frac{1}{4}\frac{\pp_\tau^2 \Omega_p^{(N)}(\tau)^2}{\Omega_p^{(N)}(\tau)^2}\,,\,\,\, \Omega_p^{(0)}(\tau)^2:=\omega_p(\tau)^2\,.\qquad
\end{eqnarray}
However, the convergence properties of this iteration are unclear. Further, there may 
be values of $p = |p|\wp$ where (\ref{iter1}) yields negative values for 
$(\Omega_p^{(N+1)})^2$ and the iteration breaks down.  
We show in this section that the iteration \eqref{iter1} is indeed well-defined  for large $|p|>0$, and obtain uniform large $|p|$ bounds for $\Omega_p^{(N)}(\tau)$ 
as well as a concomitant series expansion. In preparation we introduce  
\begin{definition}\label{defhol}
	Let $f\in C^\infty(J\times \bbS^{d-1}\times (r,\infty))$, where $J\subseteq \bbR$ is an open interval and $r>0$. Next, using the inversion mapping $\imath:(\tau, \wp,\zeta)\mapsto (\tau, \wp, \zeta\inv)$ define $\tilde{f}:=f\circ \imath\in C^\infty(J\times \bbS^{d-1}\times (0,r\inv))$. Then we say that $f\in \scrA{n}$, $n\in \bbN_0$, iff there exists $F\in C^\infty(J\times \bbS^{d-1}\times U)$, where $U\ni 0$ is open in $\bbC$, such that
\begin{enumerate}[leftmargin=8mm, rightmargin=-0mm, label=(\roman*)]
  \item $F(\tau, \wp,\zeta)=\zeta^{n}G(\tau, \wp,\zeta)$ with $G(\tau,\wp,\cdot)$ holomorphic in $U$.
  \item Restricting to $\zeta\in (0,r\inv)\cap U$, for every $\tau\in J,\,\wp\in \bbS^{d-1}$ we have 
\begin{align}
	F(\tau,\wp,\zeta)=\tilde{f}(\tau,\wp,\zeta)\,.
\end{align}
\end{enumerate}
\end{definition}
One may think of $\scrA{n}$ as consisting of functions $f(\tau,\wp,|p|)$ that extend to holomorphic ones in a neighborhood of $|p|=\infty$, with $n$ indicating the leading behavior in $|p|$ as $|p|\to\infty$. Indeed, it is clear from Definition \ref{defhol} that if $f\in \scrA{n}$ for some $n\in \bbN_0$, then all its time derivatives are uniformly bounded in $|p|$ on any compact time interval $I\subseteq J$. Precisely, for all $\ell\in \bbN_0$, there is $c_\ell (I),\,R_\ell(I)>0$ such that 
\begin{align}\label{iter2}
	\sup_{\wp\in \bbS^{d-1}}\sup_{\tau\in I}\big|\pp_\tau^\ell f(\tau, \wp, |p|)\big|\leq c_\ell(I)\mdp^{-n}\,,\quad \mdp\geq R_\ell (I)\,.
\end{align}
For the rest of this section, we fix  an arbitrary, finite and open time interval $J$ that is bounded away from the Big Bang. The main result of this section is  the following characterization of the adiabatic iterates $\Omega_p^{(N)}(\tau)$.
\begin{theorem} \label{thadmain}
	There exists a sequence of smooth functions $O_n(\tau,\wp)$, $n\in \bbN$, such that for every $N\in \bbN_0$
\begin{align}\label{iter01}
	\Omega_p^{(N)}(\tau)^2\isa \mdp^2\omega_2(\tau,\wp)^2\Big[1+\sum_{n=1}^N O_n(\tau,\wp)\mdp^{-2n}+\alpha_{N+1}(\tau,\wp,\mdp)\Big]\,,
\end{align}
with $\alpha_{N+1}\in \scrA{2N+2}$ and the implicit ``r''  large enough so that the RHS is positive. Further, $\alpha_{N+1}\circ \imath$ is an even function of $\zeta$.\footnote{Here, as well as  in the proofs of Theorem \ref{thadmain} and  Lemma \ref{lmeps}, the statement ``$f\circ\imath$ is an even function of $\zeta$'' for $f\in \scrA{n}$, $n\in \bbN_0$, means that there is an open ball $B\ni 0$ contained in  $\bbC$ such that $(f\circ\imath)(\tau,\wp,\zeta)=(f\circ\imath)(\tau,\wp,-\zeta)$ for all $\zeta\in B\,,\tau\in J,\,\wp\in \bbS^{d-1}$.} 
\end{theorem}
{\bfseries Remarks.} Before turning to the proof we note some important corollaries.

(i) Since $\alpha_{N+1}\in \scrA{2N+2}$, it admits a Taylor series expansion 
\begin{align}
	\alpha_{N+1}(\tau,\wp,\mdp)=\sum_{n=N+1}^\infty O_n^{(N+1)}(\tau,\wp)|p|^{-2n}\,,
\end{align}
where  only even powers of $\mdp$ occur since $\alpha_{N+1}\circ \imath$ is an even function of $\zeta$. Hence, for each $N\in \bbN_0$, the $(\Omega_p^{(N)})^2$ have a Taylor expansion
\begin{align}\label{iter01a}
	\Omega_p^{(N)}(\tau)^2\isa \mdp^2\omega_2(\tau,\wp)^2\Big[1+\sum_{n=1}^N O_n(\tau,\wp)\mdp^{-2n}+\sum_{n=N+1}^\infty O_n^{(N+1)}(\tau,\wp)|p|^{-2n}\Big]\,.
\end{align}
The Taylor expansion of the adiabatic iterates thus contains intrinsic information about the underlying differential equation \eqref{ad2} in the form of the $O_1,\ldots , O_N$ functions, as well as spurious parts, parameterized by the $O_n^{(N+1)}$, that change with increasing adiabatic order.

(ii) For every $N\in \bbN_0$ it is clear that the RHS of \eqref{iter01} is positive for  sufficiently large $\mdp$. Taking a square-root yields a sequence of functions $\Omega_n(\tau,\wp)$, $n\in \bbN$, such that
\begin{align}\label{iter01b}
	\Omega_p^{(N)}(\tau) = \om_2(\tau,\wp) |p| \Big\{ 1 + \sum_{n = 1}^{N} \Omega_n(\tau,\wp)
|p|^{-2n} +\beta_{N+1}(\tau,\wp,\mdp) \Big\}\,,
\end{align}
with $\beta_{N+1}\in \scrA{2N+2}$.  The $\Omega_1,\ldots, \Omega_N$ are in one-to-one 
correspondence to the $O_1,\ldots,O_N$ via
\begin{align}\label{iter01c}
	O_n(\tau,\wp)=2\Omega_n(\tau,\wp)+\sum_{j=1}^{n-1}\Omega_j(\tau,\wp)\Omega_{n-j}(\tau,\wp)\,,\quad\,n\in \bbN\,.
\end{align}
Inversely, $\Omega_1=\frac{1}{2} O_1 $, $\Omega_2=\frac{1}{2}O_2-\frac{1}{8}O_1^2$,
etc.. 

(iii) It follows readily from \eqref{iter01b} that for any compact interval $I\subseteq J$, and any $N\in \bbN$, there is a $R_N>0$ and  $0<c_N\leq d_N$ such that when $\mdp\geq R_N$
\begin{align}\label{iter01d}
	c_N(1+\mdp)\leq \Omega_p^{(N)}(\tau)\leq d_N(1+\mdp)\,,\quad \tau\in I\,,\,\wp\in \bbS^{d-1}\,.
\end{align}
Similarly, it follows that for each $\ell\in \bbN$, there are $R_{N,\ell}\geq 0$ and $c_{N,\ell}\geq 0$ such that  
\begin{align}\label{iter01e}
	\sup_{\wp\in \bbS^{d-1}}\sup_{\tau\in I}\big|\pp_\tau^\ell \Omega_p^{(N)}(\tau)\big|\leq c_\ell(I)\mdp\,,\quad \mdp\geq R_{N,\ell} (I)\,.
\end{align}

(iv) One may reexpress \eqref{iter01b} to obtain a sequence of functions $\bar{G}_n(\tau,\wp)$ such that
\be
\label{iter28} 
\frac{1}{2 \Omega_p^{(N)}(\tau)} = \frac{1}{2 \om_2(\tau,\wp) |p|} 
\Big\{ 1 + \sum_{n =1}^N (-)^n \bar{G}_n(\tau,\wp) |p|^{-2n}  + 
O( |p|^{-2N -2})\Big\} \,.
\ee 
Here, the coefficients $\bar{G}_n$ are finite combinations of the 
 $\Omega_1, \ldots, \Omega_N$, and hence are also 
robust (i.e.~unchanged at higher adiabatic order $N$). Since (\ref{ad2}) is equivalent to the Gelfand-Dickey equation 
(\ref{GD1}) for $1/(2 \Omega_p)$ (see (\ref{approx4})) the coefficients 
$\bar{G}_1, \ldots, \bar{G}_N$ will satisfy the same recursion 
relation (\ref{GD3}) (with $w = \om_2^2, v = \om_0^2$) as the GD coefficients 
$G_n$ in (\ref{GD2}). The initial values $\bar{G}_0 = G_0 =1$ coincide 
as well, so 
\be 
\label{iter29} 
\bar{G}_n = G_n, \quad n =1,\ldots , N\,.
\ee
Using {\tt Mathematica} this can also be verified explicitly to low orders.   
While the adiabatic iterates are well suited to derive the strict bounds presented in this section, they are a very roundabout conduit to extract the 
$\bar{G}_n$, which (since they coincide with the GD coefficients) are determined by a simple closed recursive system.

\medskip
We now turn to the proof of Theorem \ref{thadmain} and anticipate the  following lemma.
\begin{lemma}
	\label{lmeps}
	Define $\epsilon_n(\tau,\wp,\mdp)$,  $n\in \bbN$ by (\ref{iter1}) and  
\begin{align}\label{iter02}
	\Omega_p^{(n)}(\tau)^2=\Omega_p^{(n-1)}(\tau)^2\big(1+\epsilon_{n}(\tau,\wp,\mdp)\big)\,.
\end{align}
Then $\epsilon_n\in \scrA{2n}$.
\end{lemma}

{\itshape Proof of Theorem \ref{thadmain}:} By induction on $N\in \bbN_0$. 
For $N=0$ one has 
\begin{align}
	\Omega_p^{(0)}(\tau)^2=\omega_p(\tau)^2=\mdp^2\omega_2(\tau,\wp)^2\Big[1+\frac{\omega_0(\tau)^2}{\omega_2(\tau,\wp)^2}\mdp^{-2}\Big]\,,
\end{align}
hence here $\alpha_1(\tau,\wp,\mdp):=\omega_0(\tau)^2/[\omega_2(\tau,\wp)^2\mdp^2]$.
Since the closure of $J$ is compact, there are constants $0<c_-\leq c_+<\infty$ such that 
\begin{align}\label{iter03}
	c_-\leq \omega_2(\tau,\wp)^2\leq c_+\,,\quad \wp \in \bbS^{d-1}\,, \;\;\tau \in J\,.
\end{align}
see \eqref{om2bounds}. Consider $\tilde{\alpha}_1:=\alpha \circ \imath$, with $\imath$ the inversion map from Definition \ref{defhol};  explicitly
\begin{align}
	\tilde{\alpha}_1(\tau,\wp,\zeta)=\frac{\omega_0(\tau)^2}{\omega_2(\tau,\wp)^2}\zeta^2\,.
\end{align}
It is clear from the smoothness of $\omega_0(\tau),\,
\omega_2(\tau,\wp)$, together with \eqref{iter03}, that $\tilde{\alpha}_1$ extends to a smooth function on $J\times \bbS^{d-1}\times \bbC$ which is holomorphic in $\zeta$. Hence $\alpha_1\in \scrA{2}$, establishing the assertion for $N=0$. 

For the $N-1 \mapsto N$ induction step assume  there exist smooth functions $O_1(\tau,\wp)\,,\ldots$, $O_{N-1}(\tau,\wp)$, and $\alpha_N\in \scrA{2n}$ with $\alpha_N\circ \imath$  an even function of $\zeta$ (in some open ball around $0\in \bbC$), such that 
\begin{align}\label{iter04}
	\Omega_p^{(N-1)}(\tau)^2\isa \mdp^2\omega_2(\tau,\wp)^2\Big[1+\sum_{n=1}^{N-1} O_n(\tau,\wp)\mdp^{-2n}+\alpha_{N}(\tau,\wp,\mdp)\Big]\,.
\end{align} 
Recall that  $\Omega_p^{(N)}(\tau)^2=\Omega_p^{(N-1)}(\tau)^2(1+\epsilon_{N}(\tau,\wp,\mdp))$. Next, note $\alpha_N\,,\epsilon_N\in \scrA{2N}$, following respectively from the inductive hypothesis and Lemma \ref{lmeps}, and further $\alpha_N\circ \imath\,,\epsilon_N\circ \imath$  are both even functions of $\zeta$. We may write 
\begin{align}
	\alpha_N(\tau, \wp,\mdp)\isa O_N^{(N)}(\tau,\wp)\mdp^{-2N}+a_N(\tau,\wp,\mdp)\,,
	\nonum 
	\epsilon_N(\tau, \wp,\mdp)\isa E_N^{(N)}(\tau,\wp)\mdp^{-2N}+e_N(\tau,\wp,\mdp)\,,
\end{align}
with $a_N,\,e_N\in \scrA{2N+2}$ such that $a_N\circ\imath$ and $e_N\circ \imath$ are even functions of $\zeta$. Combined with (\ref{iter04}) it follows that
\begin{align}
	\Omega_p^{(N)}(\tau)^2\isa \mdp^2\omega_2(\tau,\wp)^2\Big[1+\sum_{n=1}^{N-1} O_n(\tau,\wp)\mdp^{-2n}+ \big(O_N^{(N)}+E_N^{(N)}\big)(\tau,\wp)\mdp^{-2N} 
	\nonum 
	&+\alpha_{N+1}(\tau,\wp,\mdp)\Big]\,,
\end{align}
with $\alpha_{N+1}$ built from products of  $a_N,\,e_N$ with 
$O_1,\ldots,O_{N-1},A_N,E_N$ terms. Thus $\alpha_{N+1}\in \scrA{2n+2}$ and we may set $O_N(\tau,\wp):=(O_N^{(N)}+E_N^{(N)})(\tau,\wp)$. Indeed, re-running through 
the argument at the next higher order one sees that $O_N$ will not be affected 
and thus defines the robust $O(|p|^{-2N})$ coefficient. Its smoothness is inherited.  
\qed

It remains to prove Lemma \ref{lmeps}.

{\itshape Proof of Lemma \ref{lmeps}:} By induction on $n\in \bbN$. 
For $n=1$ one has from (\ref{iter1}) and (\ref{iter02}) 
\begin{align}
	\epsilon_1\isa \omega^{-2}\Big(\frac{1}{16}\big(\pp_\tau \ln \omega^2\big)-\frac{1}{4}\pp_\tau^2\ln \omega^2\Big)\,,
\end{align}
where we omit the function arguments for brevity of notation. Defining $\tilde{\epsilon}_1:=\epsilon_1\circ \imath$, with $\imath$ the inversion map of Definition \ref{defhol}, gives explicitly
\begin{align}
	\tilde{\epsilon}_1(\tau,\wp,\zeta)\isa \frac{\zeta^2}{\omega_0(\tau)^2\zeta^2 + \omega_2(\tau,\wp)^2} \bigg(\frac{1}{16}\big[\pp_\tau\ln (\omega_0(\tau)^2\zeta^2 + \omega_2(\tau,\wp)^2)\big]^2
	\nonum
	&- \frac{1}{4}\pp_\tau^2\ln (\omega_0(\tau)^2\zeta^2 + \omega_2(\tau,\wp)^2)\bigg)\,.
\end{align}
Since $\omega_2(\tau,\wp)^2$ is by (\ref{iter03}) bounded away from zero on $J$, it follows that $\tilde{\epsilon}_1$ extends to a smooth function on $J\times \bbS^{d-1}\times U$ for an open set $0\in U\subseteq \bbC$ on which $\tilde{\epsilon}_1(\tau,\wp,\cdot)$ is holomorphic. Thus $\epsilon_1\in \scrA{2}$, with $\tilde{\epsilon}_1$ even in $\zeta$, as desired. 

Proceeding to the inductive step, fix $n\in \bbN$ and assume that for every $j\in \set{1,\ldots,n}$ we have $\epsilon_j\in \scrA{2j}$ with $\epsilon_j\circ\imath$ even in $\zeta$. It is not hard to show that one may write (c.f. \cite{Luders} equation (3.9))
\begin{align}\label{iter05}
	\epsilon_{n+1}& = \frac{1}{\omega^2}\frac{1}{(1+\epsilon_1)\cdots(1+\epsilon_n)}\bigg(\frac{1}{8}\frac{\dot{\epsilon}_n}{1+\epsilon_n}\pp_\tau\ln \omega^2 +\ssum{i=1}{n-1}\frac{1}{8}\frac{\dot{\epsilon}_n}{1+\epsilon_n}\frac{\dot{\epsilon}_i}{1+\epsilon_i}
	\nonum 
	 & + \frac{5}{16} \frac{\dot{\epsilon}_n^2}{(1+\epsilon_n)^2}-\frac{1}{4}\frac{\ddot{\epsilon}_n}{1+\epsilon_n}\bigg)\,.
\end{align}
Upon considering $\tilde{\epsilon}_{n+1}:=\epsilon_{n+1}\circ \imath$, it follows from the inductive hypothesis $\epsilon_j\in \scrA{2j}$, $j\in \set{1,\ldots,n}$, that $\tilde{\epsilon}_{n+1}$ extends to a smooth function on $J\times \bbS^{d-1}\times U$ for an open set $0\in U\subseteq \bbC$, and moreover $\tilde{\epsilon}_{n+1}(\tau,\wp,\cdot)$ is holomorphic on $U$. It is clear from \eqref{iter05} that the leading contribution to $\tilde{\epsilon}_{n+1}$ as $\zeta \to 0$ comes from the terms
\begin{align}
	\frac{1}{4}\frac{\zeta^2}{\omega_0^2\zeta^2 + \omega_2^2}\frac{1}{(1+\tilde{\epsilon}_1)\cdots(1+\tilde{\epsilon}_n)^2}\Big[\frac{1}{2}\pp_\tau \tilde{\epsilon}_n\,\pp_\tau\ln (\omega_0^2\zeta^2 + \omega_2^2)-\pp_\tau^2\tilde{\epsilon}_n\Big]\,,
\end{align}
where we write $\tilde{\epsilon}_j:=\epsilon_j\circ \imath$, $j\in \set{1,\ldots,n}$, and omit the $\tau,\wp$ arguments for brevity. Since $\epsilon_n\in \scrA{2n}$ (by hypothesis), and $\omega_2(\tau,\wp)^2$ is bounded away from zero, this term is $O(\zeta^{2n+2})$ as $\zeta\to 0$. Thus, we may write 
\begin{align}
	\tilde{\epsilon}_{n+1}(\tau,\wp,\zeta)=\zeta^{2n+2} h(\tau,\wp,\zeta)\,,
\end{align}
with $h(\tau,\wp,\cdot)$ holomorphic in $U$ and even in $\zeta$. Therefore we have $\epsilon_{n+1}\in \scrA{2n+2}$, as desired, completing the inductive step and the proof.
\qed


\subsection{Adiabatic vacuum states}
\label{secAdvac}

The adiabatic iteration procedure above attempts to construct  solutions to the wave equation \eqref{ad0} guided by the intuition that the time derivatives of $\om_p(\tau)^2$ are `small' 
in some sense. Recall that the $\Omega_p^{(n)}$ are characterized by Theorem \ref{thadmain} for large $\mdp$. For smaller $\mdp$ they may not even be defined. Following \cite{Luders}, we consider in the following an arbitrary continuation to all $\mdp$, and denote it by the same symbol. 
In terms of these $\Omega_p^{(n)}$ the  {\itshape adiabatic iterates of order} $n$ are defined by
\begin{eqnarray}\label{advac1}
W^{(n)}_p(\tau):= \frac{1}{\sqrt{2 \Omega_p^{(n)}(\tau)}}\exp\Big\{\!-i\!\int_{\tau_0}^\tau \!ds \,\Omega_p^{(n)}(s)\Big\}\,,\quad n\in \bbN_0\,.
\end{eqnarray}
They are themselves not solutions of \eqref{ad0}, although they do satisfy the Wronskian normalization condition $(\partial_{\tau} W^{(n)}_p \,W^{(n)\,\ast}_p)(\tau) - (\partial_{\tau} W^{(n)\,\ast}_p\,W^{(n)}_p)(\tau) = -i$. Instead, they may be used as {\itshape initial data} to construct bona-fide solutions, called {\itshape adiabatic vacua of order $n$}.
\begin{theorem}\label{thad1}
	For each $n\in \bbN_0$, the system  \eqref{ad0} admits an exact solution $S_p^{(n)}(\tau)$ such that 
	\begin{eqnarray}\label{advac1a}
	S_p^{(n)}(\tau)\is W_p^{(n)}(\tau)\big[1+O(\mdp^{-2n})\big]\,,
	\nonum 
	\pp_\tau S_p^{(n)}(\tau)\is \pp_\tau W_p^{(n)}(\tau)\big[1+O(\mdp^{-2n})\big]\,,
	\end{eqnarray}
	uniformly in $\tau\in I$ as $\mdp\to \infty$.
\end{theorem}
The proof is a straightforward adaptation of the one in \cite{Luders}; we 
include it here to cover the proof of Proposition \ref{SNviaapprox}. 
An immediate corollary is the following useful fact.
\begin{corollary} \label{corad1}
	For each $n\in \bbN_0$, the adiabatic vacua $S_p^{(n)}(\tau)$ of order $n$ have the following asymptotic behavior
	\begin{eqnarray}
	S_p^{(n)}(\tau)=O(\mdp^{-1/2})\,,\quad \forall\ell\geq 1:\,\pp_\tau^\ell S_p^{(n)}(\tau)=O(\mdp^{\ell-1/2})\,,
	\end{eqnarray}
	uniformly in $\tau\in I$.
\end{corollary}
{\itshape Proof of Corollary \ref{corad1}:} The bounds \eqref{iter01d}, \eqref{iter01e} 
entail the large $\mdp$ asymptotic behavior  $W_p^{(n)}(\tau)=O(\mdp^{-1/2})$ and $\pp_\tau W_p^{(n)}(\tau)=O(\mdp^{1/2})$, uniformly over the compact time interval $I$. By Theorem \ref{thad1} the same estimates hold for $S_p^{(n)}(\tau)$. The estimates for the higher derivatives $\pp^\ell_\tau S_p^{(n)}(\tau)$, $\ell\geq 2$,  follow from the differential equation \eqref{ad0} and the (uniform) large $\mdp$  asymptotics  of $S_p^{(n)}(\tau),\,\pp_\tau S_p^{(n)}(\tau)$.
\qed

\bigskip

{\itshape Proof of Theorem \ref{thad1}:} Fixing $n\in \bbN_0$ and substituting $\vacw{n}_p(\tau)=\itw{n}_p(\tau)\cdot R_p(\tau)$ into the differential equation $[\partial_{\tau}^2 + \omega_p(\tau)^2] \vacw{n}_p(\tau) =0$ 
yields\footnote{Although for notational brevity we omit the adiabatic order $n$, both $R_p$ and $F_p$ implicitly depend on it.}
\begin{eqnarray}\label{asym2.6}
	\partial_\tau^2 R_p+2\frac{\partial_\tau \itw{n}_p}{\itw{n}_p}\,
\partial_\tau R_p+F_p(\tau) R_p=0\,,\,\,\,F_p(\tau):=\frac{\partial_\tau^2 \itw{n}_p(\tau)+\omega_p(\tau)^2 \itw{n}_p(\tau)}{\itw{n}_p(\tau)}\,.\quad \,\,
\end{eqnarray}
It is readily verified from its definition \eqref{advac1} that $\itw{n}_p$ satisfies
\begin{eqnarray}
\bigg[\pp_\tau^2+\Omega_p^{(n)}(\tau)^2-\frac{1}{16}\bigg(\frac{\pp_\tau\Omega_p^{(n)}(\tau)^2}{\Omega_p^{(n)}(\tau)^2}\bigg)^2+\frac{1}{4}\frac{\pp_\tau^2 \Omega_p^{(n)}(\tau)^2}{\Omega_p^{(n)}(\tau)^2}\bigg]\itw{n}_p(\tau)=0\,.
\end{eqnarray}
This can be rewritten as $[ \dd_{\tau}^2 + \om_p^2] W_p^{(n)} = 
[ (\Omega_p^{(n+1)})^2-(\Omega_p^{(n)})^2] W_p^{(n)}$. Using the iteration 
equation (\ref{iter1}) this gives $F_p(\tau)=\epsilon_{n+1}(\tau,p) \Omega^{(n)}_p(\tau)^2$, with $\epsilon_j$ defined in \eqref{iter02}. The large $\mdp$ asymptotics of of $\epsilon_{n+1},\,\Omega^{(n)}$ (Lemma \ref{lmeps} and Theorem \ref{thadmain} resp.) entail that as $\mdp\to \infty$
\begin{eqnarray}\label{advac2}
F_p(\tau)=O(\mdp^{-2n})\,,\quad \tau\in I\,.
\end{eqnarray}
Next, defining the kernel
\begin{eqnarray}
\label{asym2.8}
K_p(\tau,\tau'):=\int_{\tau'}^\tau \! d\tau''\, \itw{n}_p(\tau')^2\,
\itw{n}_p(\tau'')^{-2}\,,
\end{eqnarray}
it is clear that a  function $R_p(\tau)$ satisfying 
the integral equation 
\begin{eqnarray}
\label{asym2.9}
R_p(\tau)=1-\int_{\tau_0}^\tau \! d\tau'\,K_p(\tau,\tau')F_p(\tau')R_p(\tau')\,,
\end{eqnarray}
solves (\ref{asym2.6}) subject to the boundary conditions $R_p(\tau_0)=1,\,\pp_\tau R_p(\tau_0)=0$. For such a solution we have
\begin{eqnarray}\label{advac3}
\vacw{n}_p(\tau_0)=\itw{n}_p(\tau_0)\,,\quad \pp_\tau\vacw{n}_p(\tau_0)=\pp_\tau\itw{n}_p(\tau_0)\,,
\end{eqnarray}
i.e. the adiabatic iterates provide the initial data for the construction of the adiabatic vacuum states. 
Further, 
$K_p=O(\mdp^0)$ uniformly on $I^2$; hence \eqref{advac2} entails that for sufficiently large $\mdp$ the mapping
\begin{eqnarray}
\label{asym2.10}
u(\tau)\mapsto 1-\int_{\tau_0}^\tau \! d\tau'\,K_p(\tau,\tau')F_p(\tau')u(\tau')\,,
\end{eqnarray}
is a contraction on the Banach space $\big(C([\tau_i,\tau_f],\mathbb{C}),
\left\lVert \cdot \right\rVert_{\sup}\big)$.  Thus, the integral equation \eqref{asym2.9} has a unique solution $R_p(\tau)$ by the Banach Fixed Point Theorem. Moreover, this solution may be reached (in the $\left\lVert \cdot \right\rVert_{\sup}$-norm) from the function $u\equiv1$ by repeated applications of the contractive mapping \eqref{asym2.10}. Hence, 
\begin{align}\label{asym2.11}
	R_p(\tau)=1+O(\mdp^{-2n})\,,\quad \pp_\tau R_p(\tau)=O(\mdp^{-2n})\,,
\end{align}
uniformly in $\tau\in I\,,\wp\in \bbS^{d-1}$ 
	as $\mdp\to \infty$, with the latter result following by differentiating \eqref{asym2.9}. 
In summary, we have established the existence of an exact solution $\vacw{n}_p(\tau)=\itw{n}_p(\tau)\cdot R_p(\tau)$ of \eqref{ad0}, i.e., an adiabatic vacuum state of order $n\in \bbN_0$, whose large $\mdp$ asymptotic behavior \eqref{advac1a} follows immediately from \eqref{asym2.11}. Finally, the initial condition \eqref{advac3} entails that $(\partial_{\tau} S^{(n)}_p \,S^{(n)\,\ast}_p)(\tau_0) - (\partial_{\tau} S^{(n)\,\ast}_p\,S^{(n)}_p)(\tau_0) = -i$, and since $\vacw{n}_p$ solves $[\pp_\tau^2+\omega_p(\tau)^2]\vacw{n}_p(\tau)=0$, it is Wronskian normalized for all times.
\qed

\newpage


\end{document}